\documentclass[11pt]{article}
\usepackage{amstext,amssymb,amsmath,amsbsy}

\usepackage{hyperref}
\usepackage{amscd,mathrsfs,dsfont}
\usepackage{amsfonts}
\usepackage{indentfirst}
\usepackage{verbatim}
\usepackage{amsmath}
\usepackage{amsthm}
\usepackage{enumerate}
\usepackage{graphicx}
\usepackage{subfig}
\usepackage{color}
\usepackage[OT1]{fontenc}
\usepackage[latin1]{inputenc}
\usepackage[english]{babel}
\usepackage{amssymb}
\newtheorem{theorem}{Theorem}
\newtheorem{lemma}{Lemma}

\newtheorem{proposition}{Proposition}

\newtheorem{remark}{Remark}

\newtheorem{definition}{Definition}
\numberwithin{equation}{section}

\newcommand{\proofend}{\hfill $\Box$ }

\newcommand{\dsp}{\displaystyle}

\newcommand{\bE}{{\bf E}}
\newcommand{\bH}{{\bf H}}
\newcommand{\cE}{{\cal E}}
\newcommand{\cH}{{\cal H}}

\newcommand{\oE}{E^{(1)}}
\newcommand{\oH}{H^{(1)}}
\newcommand{\tE}{E^{(2)}}
\newcommand{\tH}{H^{(2)}}

\newcommand{\hE}{\hat E}
\newcommand{\hH}{\hat H}

\newcommand{\crE}{\mathscr{E}}
\newcommand{\crH}{\mathscr{H}}

\newcommand{\supp}{\operatorname{supp}}

\newcommand{\dive}{\operatorname{div}}
\newcommand{\curl}{\operatorname{curl}}

\newcommand{\eps}{\varepsilon}

\newcommand{\loc}{_{loc}}

\newcommand{\mN}{\mathbb{N}}

\newcommand{\mR}{\mathbb{R}}

\newcommand{\mc}{\mathrm{c}}

\newcommand{\sign}{\mbox{sign }}
\newcommand{\wc}{\rightharpoonup}

\title{Superlensing using complementary media  and reflecting complementary media for electromagnetic waves}

\author{Hoai-Minh Nguyen 
\footnote{EPFL SB MATHAA CAMA, Station 8,  CH-1015 Lausanne, hoai-minh.nguyen@epfl.ch}}

\begin{document}
\date{}

\maketitle 

\begin{abstract}

In this paper, we present the proof of superlensing an arbitrary object using complementary media and  we 
study reflecting complementary media for electromagnetic waves. The analysis  is based on 
 the reflecting technique  and new results on the compactness, existence, and stability for the Maxwell equations with low regularity data.

\end{abstract}



\section{Introduction}

Negative index materials (NIMs) were first  investigated theoretically by Veselago in \cite{Veselago}. 
The  existence of NIMs was confirmed experimentally by Shelby, Smith, and Schultz in \cite{ShelbySmithSchultz}. The study of NIMs has attracted a lot of attention in the scientific community thanks to their interesting properties and many possible  applications such as  superlensing using complementary media, see \cite{NicoroviciMcPhedranMilton94, PendryNegative, PendryCylindricalLenses, PendryRamakrishna, Ng-Superlensing}, cloaking  
using complementary media, see \cite{LaiChenZhangChanComplementary, Ng-Negative-cloaking, MinhLoc2}, cloaking via 
anomalous localized resonance, see  \cite{MiltonNicorovici, BouchitteSchweizer10, AmmariCiraoloKangLeeMilton, KohnLu, MinhLoc1, Ng-CALR-CRAS, Ng-CALR, Ng-CALR-finite} and references therein.  A survey for recent mathematics progress on these applications can be found in \cite{Ng-Negative-Review}. 
In this paper, we present the proof of  superlensing using complementary media  for electromagnetic waves.

 Superlensing using  complementary media   was suggested by Veselago in \cite{Veselago} for a slab lens (a slab of index $-1$) using the ray theory.   Later, cylindrical lenses in the two dimensional  quasistatic regime, the Veselago slab  and cylindrical lenses in the finite frequency regime, and   spherical lenses in the finite frequency regime were studied by Nicorovici, McPhedran, and Milton  in \cite{NicoroviciMcPhedranMilton94},  Pendry in \cite{PendryNegative, PendryCylindricalLenses}, and  Pendry and Ramakrishna in \cite{PendryRamakrishna} respectively for constant  isotropic objects. 
Superlensing  arbitrary inhomogeneous objects using complementary media in the acoustic  setting  was established in  \cite{Ng-Superlensing} for schemes inspired from  \cite{NicoroviciMcPhedranMilton94, PendryCylindricalLenses, PendryRamakrishna} and guided by  the concept of reflecting complementary media  in \cite{Ng-Complementary}. The proof of  superlensing  arbitrary inhomogeneous objects  using complementary media for electromagnetic waves presented in this paper therefore represents the natural completion of this line of work. 

\begin{figure}[h!]
\begin{center}
\includegraphics[width=10cm]{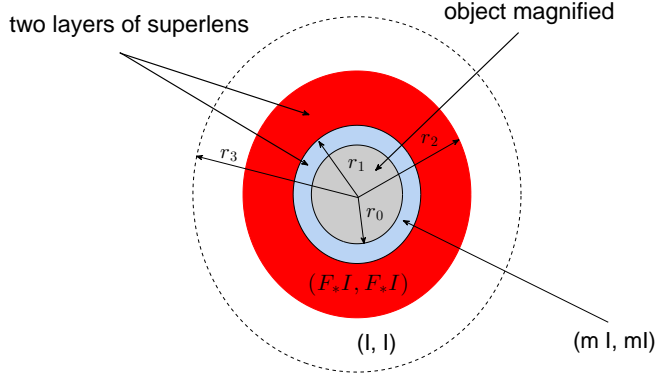}
\caption{Lens contains two layers: the outer layer using NIMs is  given by \eqref{first-part}, the inner layer  is given by \eqref{second-part}}.  \label{fig1}
\end{center}
\end{figure}

Let us describe how to magnify $m$ times (m is a given real number greater than  1) the region $B_{r_0}$ for some $r_0> 0$ in which the medium is characterized by a pair of two uniformly elliptic matrix-valued functions $(\eps_O, \mu_O)$  using complementary media.  The idea suggested by Pendry and Ramakrishna in  \cite{PendryRamakrishna} is to put a lens in $B_{r_2} \setminus B_{r_0}$ whose medium is  characterized by  $\Big(- \big(r_2^2/|x|^2 \big)I, - \big(r_2^2/|x|^2 \big)I \Big)$; the loss is ignored. Our lens construction is as follows.  Let  $\alpha, \beta > 1$ be such that 
\begin{equation}\label{ab}
\alpha \beta - \alpha - \beta = 0.
\end{equation}
Set 
\begin{equation}\label{choice-123}
r_1 = m^{1- 1/\alpha} r_0 , \quad   r_2 = m r_0, \quad \mbox{ and } \quad r_3 = m^{2 - 1/ \alpha} r_0, 
\end{equation}
and define $F: B_{r_2} \setminus \{ 0 \} \to \mR^3 \setminus \bar B_{r_2}$ and $G: \mR^3 \setminus \bar B_{r_3} \to B_{r_3} \setminus \{0\}$ by 
\begin{equation*}
 F(x) = r_2^\alpha x/ |x|^\alpha \quad \mbox{ and }  \quad G(x) = r_3^\beta x/ |x|^\beta. 
\end{equation*}
Our lens contains two parts (see Figure~\ref{fig1}). The first one of NIMs is given by 
\begin{equation}\label{first-part}
\big(F^{-1}_*I, F^{-1}_*I\big) \quad \mbox{ in } B_{r_2} \setminus B_{r_1}
\end{equation} 
(see \eqref{formula-eps-mu} below for the explicit formula)  and the second one is
\begin{equation}\label{second-part}
\big(m I, m I \big) \quad \mbox{ in } B_{r_1} \setminus B_{r_0}. 
\end{equation}

Given a diffeomorphism ${\cal T}$ from $D$ onto $D'$, the following standard notations are used 
\begin{equation}\label{def-TT}
{\cal T}_*a (x') = \frac{\nabla {\cal T}(x) a(x) \nabla {\cal T}^T(x)}{\det \nabla {\cal T}(x)}  \quad \mbox{ and } \quad  {\cal T}_*j (x')= \frac{\nabla {\cal T}(x) j(x)}{\det \nabla {\cal T}(x)},
\end{equation}
with  $x' ={\cal  T}(x)$,  for a matrix-valued function $a$ and   a vector-valued function $j$ defined in $D$.  

As showed later in Section~\ref{sect-thm-lens}, we have
\begin{equation}\label{formula-eps-mu}
F^{-1}_*I = -  \frac{r_2^\alpha}{ r^{\alpha}}  \left[ \frac{1}{\alpha - 1} e_{r} \otimes e_{r } + (\alpha -1) \Big( e_{\theta} \otimes e_{\theta}  + e_{\theta} \otimes e_{\varphi}  \Big) \right] \mbox{ in } B_{r_2} \setminus B_{r_1}. 
\end{equation}
Letting $\alpha = \beta = 2$, one rediscovers the construction suggested by Pendry and Ramakrishna in  $B_{r_2} \setminus B_{r_1}$. Note that even in this case, our lens construction contains two layers and is different from theirs where one layer is used.  We emphasize here that the lens-construction is independent of the object. 
Taking into account the loss, the medium is characterized by $(\eps_\delta, \mu_\delta) $, 
where 
\begin{equation}\label{def-epsmu}
(\eps_\delta, \mu_\delta) = \left\{ \begin{array}{cl} \big(F^{-1}_*I + i \delta I, F^{-1}_*I + i \delta I\big) & \mbox{ in } B_{r_2} \setminus B_{r_1}, \\[6pt]
\big(m I, m I \big) & \mbox{ in } B_{r_1} \setminus B_{r_0}, \\[6pt]
(\eps_O, \mu_O) & \mbox{ in } B_{r_0},\\[6pt]
(I, I)  & \mbox{ otherwise}. 
\end{array} \right. 
\end{equation}

\medskip
Some comments on the construction are necessary. The media  $(\eps_0, \mu_0)$ in $B_{r_2} \setminus B_{r_1}$ and $(I, I)$ in  $B_{r_3} \setminus B_{r_2}$ are complementary or more precisely reflecting complementary (see Section~\ref{sect-2}). 
For a given $r_2$, we choose $r_1$ and $r_3$  such that  $r_3/ r_1 = m$ and $F(\partial B_{r_1}) = \partial B_{r_3}$,  
since a superlens of  $m$ times magnification is considered. 
The choice of $(\eps_\delta, \mu_\delta)=(\eps_0, \mu_0) = (mI, mI)$ in $B_{r_1} \setminus B_{r_0}$  and $r_2 = m r_0$ is to ensure, by \eqref{ab},  that
\begin{equation}\label{the-choice}
(G_*F_* \eps_0, G_*F_* \mu_0)  = (I, I) \mbox{ in } B_{r_3} \setminus B_{r_2}.  
\end{equation}
Fix $k>0$. Given $j \in L^{2}_{\mc}(\mR^3)$ with $\supp j \subset \subset \mR^3 \setminus B_{r_3}$, let  $(E_\delta, H_\delta), (\hE, \hH) \in [H_{\loc}(\curl, \mR^3)]^2$ ($\delta > 0$) be respectively the unique outgoing solution  to 
\begin{equation}\label{sys-EH-delta}
\left\{ \begin{array}{lll}
\nabla \times E_\delta &= i k \mu_\delta H_\delta & \mbox{ in } \mR^3, \\[6pt]
\nabla \times H_\delta & = - i k \eps_\delta E_\delta +  j & \mbox{ in }  \mR^3,  
\end{array} \right.
\end{equation}
and 
\begin{equation}\label{sys-hat-EH}
\left\{ \begin{array}{lll}
\nabla \times \hE &= i k \hat  \mu \hH & \mbox{ in } \mR^3, \\[6pt]
\nabla \times \hH & = - i k \hat \eps \hE + j & \mbox{ in }  \mR^3,  
\end{array} \right.
\end{equation}
where
\begin{equation*}
(\hat \eps, \hat  \mu) = \left\{ \begin{array}{cl} (I, I) & \mbox{ in } \mR^3 \setminus  B_{m r_0}, \\[6pt]
\big(m^{-1}\eps_O(x/ m), m^{-1}\mu_O(x/ m) \big)  & \mbox{ otherwise}. 
\end{array} \right. 
\end{equation*} 
Recall that a solution $(E, H) \in [H_{\loc}(\curl, \mR^3 \setminus B_{R})]^2$ (for some $R>0$) to the system 
\begin{equation*}
\left\{ \begin{array}{lll}
\nabla \times E &= i k H & \mbox{ in } \mR^3 \setminus B_R, \\[6pt]
\nabla \times H & = - i k E & \mbox{ in }  \mR^3 \setminus B_R, 
\end{array} \right.
\end{equation*}
is said to satisfy the outgoing condition (or the Silver-M\"uller radiation condition) if 
\begin{equation}\label{OC}
E \times x + r H = O(1/r) \mbox{ as } r = |x| \to + \infty. 
\end{equation}

Our  result on superlensing is the following theorem. 

\begin{theorem}\label{thm-lens} Let  $j \in [L^{2}(\mR^3)]^3$ with $\supp j \subset B_{R_0} \setminus B_{r_{3}}$ for some $R_0>0$,  and  $(E_\delta, H_\delta),  (\hE, \hH) \in [H_{\loc}(\curl, \mR^3)]^2$ be the unique outgoing solutions to \eqref{sys-EH-delta} and \eqref{sys-hat-EH} respectively. We have, for $R>0$,  
\begin{equation*}
\|(E_\delta, H_\delta) - (\hE, \hH) \|_{H(\curl, B_R \setminus B_{r_3})} \le C_R \delta^{1/2 }\| j\|_{L^2}, 
\end{equation*}
for some positive constant $C_R$ independent of $\delta$ and $j$.  In particular,  
\begin{equation}\label{key-point}
(E_{\delta}, H_\delta) \to  (\hE, \hH) \mbox{ in } [H_{\loc} (\curl, \mR^3 \setminus B_{r_3})]^2 \mbox{ as } \delta \to 0. 
\end{equation}
\end{theorem}

For an observer outside $B_{r_3}$, the object $(\eps_O, \mu_O)$ in $B_{r_0}$ would act like 
\begin{equation*}
\big( m^{-1} \eps_O (x/m), m^{-1} \mu_O(x/m) \big) \mbox{ in } B_{m r_0}
\end{equation*}
by \eqref{key-point}: one has a superlens whose magnification is $m$.  

The proof of  
Theorem~\ref{thm-lens} given in Section~\ref{sect-thm-lens} is derived from Theorem~\ref{thm1} in Section~\ref{sect-2}.  Section~\ref{sect-2} is devoted to the concept of reflecting complementary media (Definition~\ref{def-Geo}) and their properties (Theorem~\ref{thm1}). This concept appears naturally in the study of superlensing mentioned above and is inspired from \cite{Ng-Complementary}. 
The analysis of Theorem~\ref{thm1} is based on 
 the reflecting technique which has root from \cite{Ng-Complementary} and a number of new results on the compactness, existence, and stability for the Maxwell equations with low regularity data. 

\medskip 
The paper is organised as follows. In Section~\ref{sect-2}, we discuss reflecting complementary media. Proof of Theorem~\ref{thm-lens} is given in Section~\ref{sect-thm-lens}. 




\section{Reflecting complementary media} \label{sect-2}

Let $ \Omega_{1} \subset \subset\Omega_{2}$
 be smooth simply connected bounded open subsets of $\mR^3$. Let $\eps, \mu$ be two {\bf real}  measurable matrix-valued functions defined in $\mR^3$. We assume that $\eps, \mu$ are {\bf bounded} in $\mR^3$ and uniformly elliptic in $\mR^3 \setminus (\Omega_2 \setminus \Omega_1)$, i.e., for some $1\le \Lambda < + \infty$, 
\begin{equation}\label{pro-eps-mu}
\frac{1}{\Lambda} |\xi|^2 \le \langle \eps(x) \xi, \xi \rangle \le \Lambda |\xi|^2 \mbox{  and  } \frac{1}{\Lambda} |\xi|^2 \le \langle \mu(x) \xi, \xi \rangle \le \Lambda |\xi|^2  \quad \forall \, \xi \in \mR^3, \mbox{ a.e. } x \in \mR^3 \setminus (\Omega_2 \setminus \Omega_1), 
\end{equation}
and
\begin{equation}\label{pro-Identity}
\eps = \mu = I \mbox{ in } \mR^3 \setminus B_{R_0}, 
\end{equation}
for some $R_0 > 0$ with $\Omega_2 \subset \subset B_{R_0}$. Here and in what follows, $\langle \cdot, \cdot, \rangle$ denotes the Euclidean  scalar product. 
 We also assume that \footnote{This condition is used for various uniqueness statements obtained by the unique continuation principle.}
\begin{equation}\label{eps-mu-C1}
(\eps, \mu) \mbox{ is piecewise } C^1.
\end{equation} 
Set, for $\delta \ge 0$, 
\begin{equation}\label{def-eDelta}
(\eps_\delta, \mu_\delta) = \left\{\begin{array}{cl} (\eps + i \delta I, \mu + i \delta I) & \mbox{ if }  x \in \Omega_2 \setminus \Omega_1,  \\[6pt]
(\eps, \mu) & \mbox{ otherwise}.
\end{array}\right.
\end{equation}
It is clear that $(\eps_0, \mu_0) = (\eps, \mu)$ in $\mR^3$.
Note that we do not impose the ellipticity of $\eps$ and $\mu$ in $\Omega_2 \setminus \Omega_1$. In fact, as seen later, in the setting of reflecting complementary media, they are negative (see Remark~\ref{rem-Negative}).   Fix $k>0$. Given $j \in L^2_{\mc}(\mR^3)$,  we are interested in  the behavior of the unique outgoing solution $(E_\delta, H_\delta) \in [H_{\loc}(\curl, \mR^3)]^2$ ($\delta >0$) to the Maxwell system
\begin{equation}\label{eq-EHDelta}
\left\{ \begin{array}{llll}
\nabla \times E_\delta &= & i k \mu_\delta H_\delta & \mbox{ in } \mR^3,\\[6pt]
\nabla \times H_\delta & =&  - i k \eps_\delta E_\delta + j & \mbox{ in }  \mR^3, 
\end{array} \right.
\end{equation}
as $\delta \to 0$ in the case $(\eps, \mu)$ satisfies  the reflecting complementary property, a concept introduced   in Definition~\ref{def-Geo} below. 

For  an open subset $\Omega$ of $\mR^3$,  the following standard notations  are used: 
\begin{equation*}
H(\curl, \Omega) : = \big\{ u \in [L^2(\Omega)]^3; \; \nabla \times u  \in [L^2(\Omega)]^3 \big\}, 
\end{equation*}
\begin{equation*}
\| u\|_{H(\curl, \Omega)} : = \| u\|_{L^2(\Omega)} + \| \nabla \times u \|_{L^2(\Omega)}, 
\end{equation*}
\begin{equation*}
H_{\loc}(\curl, \Omega) : = \big\{ u \in [L_{\loc}^2(\Omega)]^3; \; \nabla \times u  \in [L^2_{\loc}(\Omega)]^3 \big\}. 
\end{equation*}

We are ready to introduce

\begin{definition}[Reflecting complementary media] \label{def-Geo} Let $\Omega_1 \subset \subset \Omega_2 \subset \subset \Omega_3$ be smooth simply connected bounded open subsets of $\mR^3$. 
The media $(\eps, \mu)$ in $\Omega_2 \setminus \Omega_1$ and $(\eps, \mu)$ in $\Omega_3 \setminus \Omega_2$ are reflecting complementary if there exists a diffeomorphism $F: \Omega_2 \setminus \bar \Omega_1 \to \Omega_3 \setminus \bar \Omega_2$ such that 
\begin{equation}\label{cond-eps-mu}
(\eps, \mu) = (F_*\eps, F_*\mu)  \mbox{ in } \Omega_3 \setminus \Omega_2, 
\end{equation} 
\begin{equation}\label{cond-F-boundary}
F(x) = x \mbox{ on } \partial \Omega_2,  
\end{equation}
and the following two conditions hold: 1) There exists a diffeomorphism extension of $F$, which is still denoted by  $F$, from $\Omega_{2} \setminus \{x_{1}\}$ onto $\mR^3 \setminus \bar \Omega_{2}$ for some $x_{1} \in \Omega_{1}$.  2) There exists a diffeomorphism $G: \mR^3 \setminus \bar \Omega_{3} \to \Omega_{3} \setminus \{x_{1} \}$  such that  $G \in C^1(\mR^3 \setminus \Omega_3)$,  $G(x) = x \mbox{ on } \partial \Omega_3$,
and $G \circ F : \Omega_1  \to \Omega_3 \mbox{ is a diffeomorphism if one sets } G\circ F(x_1) = x_1$. 
\end{definition}

Here and in what follows, when we mention a diffeomorphism $F:  \Omega \to \Omega'$ for two open {\bf smooth} subsets $\Omega, \Omega'$ of $\mR^d$, we mean that $F$ is a diffeomorphism, $F \in C^1 (\bar \Omega)$, and $F^{-1} \in C^1(\bar \Omega')$.

\begin{figure}[h!]
\begin{center}
\hspace{-1cm}\includegraphics[width=10cm]{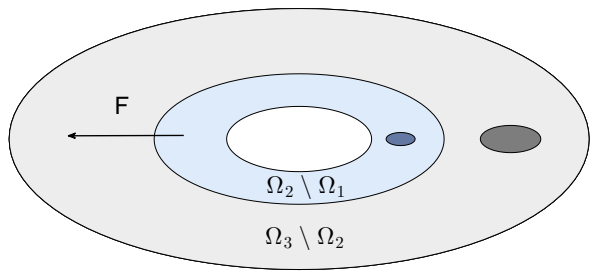} 
\caption{The media $(\eps, \mu)$ in $\Omega_2 \setminus \Omega_1$ and $(\eps, \mu)$ in $\Omega_3 \setminus \Omega_2$ are reflecting complementary media if roughly speaking  $(F_*\eps, F_*\mu) = (\eps, \mu)$ in $\Omega_3 \setminus \Omega_2$ for some  diffeomorphism  $F$  from $\Omega_2 \setminus  \bar \Omega_1$ to $\Omega_3 \setminus \bar \Omega_2$ such that $F(x) = x$ on $\partial \Omega_2$} \label{fig-C}
\end{center}
\end{figure}

The illustration of reflecting complementary media is given in Figure ~\ref{fig-C}. Note that the superlensing setting in Theorem~\ref{thm-lens} has this property. Theorem~\ref{thm-lens} will be derived from Theorem~\ref{thm1}-below, on properties of the reflecting complementary media. 

\begin{remark} \fontfamily{m} \selectfont We emphasize here that in \eqref{def-TT} $\det DT(x)$ is used not $|\det DT(x)|$
and  in \eqref{cond-eps-mu}, one requires that  $(\eps, \mu) = (F_*\eps, F_*\mu) $ not 
$(\eps, \mu) = (-F_*\eps, -F_*\mu)$. These conventions are different from the ones in the acoustic setting, see,  e.g.,   \cite{Ng-Complementary}, and  are more convenient in the study of  Maxwell equations. 
\end{remark}

\begin{remark} \fontfamily{m} \selectfont \label{rem-Negative} Assume that $(\eps, \mu)$ in $\Omega_2 \setminus \Omega_1$ and $(\eps, \mu)$ in $\Omega_3 \setminus \Omega_2$ are reflecting complementary and $(\eps, \mu)$ is positive in $\Omega_3 \setminus \Omega_2$. Since $F(x) = x $ on $\partial \Omega_2$ and $F: \Omega_2 \setminus \Omega_1 \to \Omega_3 \setminus \Omega_2$ is a diffeomorphism, it follows that $\det \nabla F(x) < 0$ in $\Omega_2 \setminus \Omega_1$. Therefore, 
$(\eps, \mu)$ is negative in $\Omega_2 \setminus \Omega_1$. 
\end{remark}

We next make some comments on the definition. Condition \eqref{cond-eps-mu} implies that $(\eps, \mu)$ in $\Omega_{2} \setminus \Omega_{1}$ and $(\eps, \mu)$ in $\Omega_{3} \setminus \Omega_{2}$ are complementary in the ``usual" sense \footnote{In fact, the concept of complementary media have not be defined in a precise manner. Property \eqref{cond-eps-mu} mentioned here is the common point in some examples discussed in the literature.}. The term ``reflecting'' in the definition comes from \eqref{cond-F-boundary} and the assumption $\Omega_{1} \subset \Omega_{2} \subset \Omega_{3}$. Conditions  \eqref{cond-eps-mu} and \eqref{cond-F-boundary} are the main assumptions in the definition. They are motivated by the following observation. Assume that $(\eps, \mu)$ in $\Omega_2 \setminus \Omega_1$ and $(\eps, \mu)$ in $\Omega_3 \setminus \Omega_2$ are reflecting complementary and suppose that there exists a solution $(E_0, H_0) \in [H(\curl , \Omega_3 \setminus \Omega_1)]^2$  of 
\begin{equation}\label{eq-EHDelta}
\left\{ \begin{array}{llll}
\nabla \times E_0 &= & i k \mu H_0 & \mbox{ in } \Omega_3 \setminus \Omega_1,\\[6pt]
\nabla \times H_0 & =&  - i k \eps E_0 & \mbox{ in }  \Omega_3 \setminus \Omega_1. 
\end{array} \right.
\end{equation}
For $x' \in \Omega_3 \setminus \Omega_2$, define $(\oE_{0}(x'), \oH_{0}(x')) =\big(\nabla F^{-T}(x) E_0 (x) ,  \nabla F^{-T}(x) H_0 (x) \big)$ where $x = F^{-1}(x')$.  Conditions  \eqref{cond-eps-mu} and \eqref{cond-F-boundary} imply that, by the rule of change of variables (see e.g. Lemma~\ref{lem-CV} in Section~\ref{sect-pre}), 
\begin{equation}\label{eq-EHDelta}
\left\{ \begin{array}{llll}
\nabla \times (E_0 - \oE_0)&= & i k \mu (H_0 - \oH_0) & \mbox{ in } \Omega_3 \setminus \Omega_2,\\[6pt]
\nabla \times (H_0 - \oH_0) & =&  - i k \eps (E_0 - \oE_0) & \mbox{ in }  \Omega_3 \setminus \Omega_2, \\[6pt] 
(E_0 - \oE_0) \times \nu & = &  (E_0 - \oE_0) \times \nu = 0 &  \mbox{ on } \partial \Omega_2. 
\end{array} \right.
\end{equation}
 Hence, $(E_0, H_0) = (\oE_{0}, \oH_{0})$ in $\Omega_3 \setminus \Omega_2$  if $(\eps, \mu)$  is uniformly elliptic in $\Omega_3 \setminus \Omega_2$ by the unique continuation principle;  this is the main motivation for conditions~\eqref{cond-eps-mu} and \eqref{cond-F-boundary}.  Conditions 1) and 2) are  mild assumptions. Introducing $G$ in the definition makes the analysis more accessible; see  Sections~\ref{sect-complementary}. 

\medskip

Here and in what follows, we denote 
\begin{equation}\label{def-hat-eps-mu}
(\hat \eps, \hat \mu) := \left\{ \begin{array}{cl}
(\eps, \mu) & \mbox{ if } x \in \mR^3 \setminus \Omega_3,\\[6pt]
(G_*F_*\eps, G_*F_*\mu) & \mbox{ if } x \in \Omega_3. 
\end{array}\right.
\end{equation}

The following definition is used in the statement of Theorem~\ref{thm1} below.

\begin{definition}[Compatibility condition]\label{compatible-k}Assume that $(\eps, \mu)$ in $\Omega_2 \setminus \Omega_1$ and $(\eps, \mu)$ in $\Omega_3 \setminus \Omega_2$ are reflecting complementary for some $\Omega_2 \subset \subset \Omega_3 \subset \subset \mR^3$.   Then $j \in [L^2_{\mc}(\mR^3)]^3$ {\bf with} $\supp j \cap \Omega_3 = \O$ is said to be compatible if and only if  there exists
$(\bE, \bH) \in [H(\curl, \Omega_3 \setminus \Omega_2)]^2$ such that
\begin{equation}\label{def-comp-E2H2}
\left\{\begin{array}{cllll}
\nabla \times \bE = i k  \mu \bH & \mbox{ in } \Omega_3 \setminus \Omega_2,  \\[6pt]
\nabla \times \bH  = - i k  \eps \bE  & \mbox{ in } \Omega_3 \setminus \Omega_2, \\[6pt]
\bE \times   \nu  =  \hE \times \nu, \;   \bH \times \nu = \hH \times \nu & \mbox{ on } \partial \Omega_3,  
\end{array}\right.
\end{equation}
where  $(\hE, \hH) \in [H_{\loc}(\curl, \mR^3)]^2$ is the unique outgoing solution to the system
\begin{equation}\label{bEbH}\left\{
\begin{array}{clll}
\nabla \times \hE  =  i k \hat \mu \hH & \mbox{ in } \mR^3, \\[6pt]
\nabla \times \hH  =  - i k \hat \eps  \hE +j & \mbox{ in }  \mR^3. 
\end{array}\right. 
\end{equation}
\end{definition}

\begin{remark}  \fontfamily{m} \selectfont  It is important to note that $\hat \eps$ and $\hat \mu$ are uniformly elliptic in $\mR^3$ by \eqref{def-hat-eps-mu} since $\det \nabla F$ and $\det \nabla G$ are both negative. The existence and uniqueness of $(\hE,\hH)$ then follow from Lemma~\ref{lem-stability1} in Section~\ref{sect-pre}. The uniqueness of   $(\bE, \bH)$ is a consequence of the unique continuation principle (see \cite{Tu, BallCapdeboscq}). 
\end{remark}

\begin{remark} \label{rem-Cauchy}  \fontfamily{m} \selectfont Note that \eqref{def-comp-E2H2} is a Cauchy problem:  the uniqueness is ensured by the unique continuation principle but the existence is not; hence the resonance might appear. 
\end{remark}

Our main result on the reflecting complementary media for electromagnetic waves is:

\begin{theorem}\label{thm1} Let $k>0$, $0< \delta < 1$, $j \in [L^2(\mR^3)]^3$ with compact support,  and let $(E_\delta, H_\delta) \in [H_{\loc}(\curl, \mR^3)]^2$ be the unique outgoing solution of \eqref{eq-EHDelta}. 
Assume that   $(\eps, \mu)$ in $\Omega_{2} \setminus \Omega_{1}$ and $(\eps, \mu)$ in $\Omega_{3} \setminus \Omega_{2}$  are  reflecting complementary for some $\Omega_2 \subset \subset \Omega_3 \subset \subset \mR^3$  and $\supp j \cap \Omega_3 = \O$. We have
\begin{enumerate}
\item[a)] Case 1: $j$ is compatible. There exists a unique outgoing solution  $(E_0, H_0) \in [H_{\loc}(\curl, \mR^3)]^2$ to
\begin{equation}\label{limit-EH-0}
\left\{ \begin{array}{lll}
\nabla \times E_0 &= i k  \mu H_0 & \mbox{ in } \mR^3, \\[6pt]
\nabla \times H_0 & = - i k  \eps E_0 + j & \mbox{ in }  \mR^3. 
\end{array} \right.
\end{equation}
Moreover, 
$$
(E_0, H_0) = (\hE, \hH) \mbox{ in } \mR^3 \setminus \Omega_3, 
$$
and, for all $R>0$, 
$$
\| (E_\delta, H_\delta) - (E_0, H_0) \|_{H(\curl, B_R)} \le C_R \delta^{1/2} \| (E_0, H_0)\|_{L^2\big( (\Omega_2 \setminus \Omega_1) \cup B_R\big)},  
$$
for some positive constant $C_R$ independent of $j$ and $\delta$.

\item[b)] Case 2: $j$ is not compatible. We have
\begin{equation}\label{limit-energy-k}
\lim_{\delta \to 0}\| (E_\delta, H_\delta) \|_{H(\curl, B_{R})} = + \infty, 
\end{equation}
for $R > 0$  such that $\bar \Omega_2  \subset B_{R}$. 
\end{enumerate}
\end{theorem}

The implication of Theorem~\ref{thm-lens} from Theorem~\ref{thm1} is given in Section~\ref{sect-thm-lens}. 

\medskip 

The rest of this section  containing two subsections is devoted to the proof of Theorem~\ref{thm1}. In the first one, we presents some lemmas used in the proof of Theorem~\ref{thm1}. The proof of Theorem~\ref{thm1} is given in the second subsection.

\subsection{Some useful lemmas}\label{sect-pre}

In this section, we present some technical lemmas which are used in the proof of Theorem~\ref{thm1}. 
\medskip
The following compactness result plays an important role in our analysis. 
\begin{lemma} \label{lem-compactness} Let $D$ be a bounded smooth open subset of $\mR^3$, $(u^{(n)}) \subset H(\curl, D)$, and $\eps$ be a symmetric uniformly elliiptic matrix-valued function defined in $D$.  Assume that
\begin{equation*}
\sup_{n \in \mN} \| u^{(n)} \|_{H(\curl, D)} < + \infty,  
\end{equation*}
and  \footnote{$H^{-1}(D)$ denotes the duality of $H^1_0(D)$.} 
\begin{equation}\label{WC-condition}
\Big( \nabla \cdot (\eps u^{(n)}) \Big) \mbox{ converges in } H^{-1}(D) \mbox{ and } \big(u^{(n)} \times \nu  \big) \mbox{ converges in } [H^{-1/2}(\partial D)]^3. 
\end{equation}
There exists a subsequence  of $(u^{(n)}) $  which converges in $[L^2(D)]^3$. 
\end{lemma}

\noindent{\bf Proof.}  We first assume that $D$ is simply connected. Let $B$ be an open  ball such that $\bar D  \subset B$. Let $\varphi^{(n)} \in H^1(B \setminus D)$ be the unique solution with zero mean, i.e., $\int_{B \setminus D} \varphi_n = 0$,  to 
\begin{equation*}
\left\{\begin{array}{cl}
-\Delta \varphi^{(n)} = 0 & \mbox{ in } B \setminus D, \\[6pt]
\partial_\nu \varphi^{(n)} = (\nabla \times u^{(n)}) \cdot \nu & \mbox{ on } \partial D, \\[6pt]
\partial_\nu \varphi^{(n)}  = 0 & \mbox{ on } \partial B. 
\end{array} \right.  
\end{equation*} 
The existence of $\varphi^{(n)}$ is a consequence of the fact 
\begin{equation*}
\int_{\partial D} (\nabla \times u^{(n)}) \cdot \nu = 0, 
\end{equation*}
since $\nabla \cdot (\nabla \times u^{(n)}) = 0 $ in $D$. Set
\begin{equation}\label{def-chi}
\chi^{(n)} = \left\{ \begin{array}{cl} \nabla \times u^{(n)} & \mbox{ in } D, \\[6pt]
\nabla \varphi^{(n)} & \mbox{ in } B \setminus D, \\[6pt]
0 & \mbox{ in } \mR^3 \setminus B. 
\end{array}\right. 
\end{equation}
It is clear that  $\nabla \cdot \chi^{(n)} = 0$ in $ \mR^3$. Set \footnote{The notation $*$ here means the convolution.}
\begin{equation}\label{def-Phi}
 \Psi^{(n)} = G*\chi^{(n)} \mbox{ in } \mR^3,  
\end{equation}
where $G$ is the fundamental solution to the Laplace equation in $\mR^3$; this implies $-\Delta \Psi^{(n)} = \chi^{(n)}$ in $\mR^3$
and
\begin{equation}\label{H2}
\|\Psi_n \|_{H^2(B)} \le C \| \chi_n\|_{L^2}. 
\end{equation}
Here and in what follows,  $C$ denotes a positive constant depending only on $B$ and $D$. 
Since $ \nabla \cdot \chi^{(n)} = 0$ in $\mR^3$, it follows that  
\begin{equation}\label{divPsi}
\nabla \cdot \Psi^{(n)} = 0 \mbox{ in } \mR^3. 
\end{equation}
Set 
\begin{equation*}
w^{(n)} = \nabla \times \Psi^{(n)} \mbox{ in } D. 
\end{equation*}
We derive from \eqref{H2} that  $(w^{(n)})$ is bounded in $[H^1(D)]^3$.   Without loss of generality, one may assume that 
\begin{equation}\label{wn}
(w^{(n)}) \mbox{ converges in } [L^2(D)]^3.
\end{equation}  
Using the fact that 
\begin{equation*}
\nabla \times (\nabla \times  \Psi^{(n)} )= \nabla (\nabla \cdot \Psi^{(n)}) - \Delta \Psi^{(n)} \mbox{ in } D, 
\end{equation*}
we derive from   \eqref{divPsi} that 
\begin{equation*}
 \nabla \times  w^{(n)} = \nabla \times  u^{(n)} \mbox{ in } D. 
\end{equation*}
Since $D$ is simply connected, one has
\begin{equation*}
u^{(n)} = w^{(n)} + \nabla p^{(n)} \mbox{ in } D, 
\end{equation*}
for some $p^{(n)} \in H^1(D)$ such that $\int_{\partial D} p^{(n)} = 0$ (see,  e.g.,  \cite[Theorem 3.37]{Monk}); hence
\begin{equation*}
\nabla \cdot (\eps \nabla p^{(n)}) = \nabla \cdot (\eps u^{(n)}) - \nabla \cdot  (\eps w^{(n)}) \mbox{ in } D. 
\end{equation*}
A combination of  \eqref{WC-condition} and \eqref{wn} yields
\begin{equation}\label{pn}
\Big(\nabla \cdot (\eps \nabla p^{(n)}) \Big) \mbox{ converges in } H^{-1}(D).  
\end{equation}
On the other hand, 
\begin{equation*}
\|p^{(n)} - p^{(m)} \|_{H^{1/2}(\partial D)} \le C  \|\nabla p^{(n)} \times \nu - \nabla p^{(m)} \times \nu \|_{H^{-1/2}(\partial D)} 
\end{equation*}
 since $\int_{\partial D} p^{(n)} = 0$; which implies 
\begin{multline*}
\|p^{(n)} - p^{(m)} \|_{H^{1/2}(\partial D)} \\[6pt] 
\le C  \|u^{(n)} \times \nu - u^{(m)} \times \nu  \|_{H^{-1/2}(\partial D)}  + C  \|w^{(n)} \times \nu - w^{(m)} \times \nu \|_{H^{-1/2}(\partial D)}.  
\end{multline*}
Since $(w^{(n)})$ is bounded in $[H^1(D)]^3$ and converges in $[L^2(D)]^3$,   it follows  from \eqref{WC-condition} that $(p^{(n)})$ converges in $H^{1/2}(\partial D)$.  Combining this, \eqref{WC-condition},  and \eqref{pn},  we derive that $(p^{(n)})$ converges in $H^1(D)$.  Since $u^{(n)} = w^{(n)} + \nabla p^{(n)}$ and  $(w^{n})$ converges in $[L^2(D)]^3$,  we derive that $(u^{(n)})$ converges in $[L^2(D)]^3$. The  proof is complete in the case $D$ is simply connected. The proof in the general case follows by using local charts.   \proofend

\begin{remark}  \fontfamily{m} \selectfont Lemma~\ref{lem-compactness} is known if instead of \eqref{WC-condition}  one assumes  that \begin{equation*}
 \big(\nabla \cdot (\eps u^{(n)}) \big) \mbox{ is bounded in  } L^2(D) \mbox{ and } \big(u^{(n)} \times \nu \big) \mbox{ is bounded in  } [L^{2}(\partial D)]^3,  
\end{equation*}
(see \cite{Weber}). It is clear that Lemma~\ref{lem-compactness} implies the known compactness result. 
The case $\eps = I$ was established in \cite[Lemma A5]{HaddarJolyNguyen2} under the additional assumption $\Big(\nabla \cdot (\eps u^{(n)}) \Big)$ is bounded in $L^2$. The proof presented here is in the same spirit of the one given in  \cite{HaddarJolyNguyen2}, which  has roots from  \cite{Costabel}. Condition~\eqref{WC-condition} appears naturally when one studies the existence and the stability for Maxwell equations (see Lemmas~\ref{lem-trace}, \ref{lem-stability-inside}, and \ref{lem-stability1}). 
\end{remark}

The second lemma is a known result  on the trace of $H(\curl, D)$ (see \cite{AlonsoValli, BCostabel, Paquet}). 

\begin{lemma} \label{lem-trace} Let $D$ be a smooth open bounded subset of $\mR^3$ and set $\Gamma = \partial D$. The tangential trace operator 
\begin{equation*}
\begin{array}{cccc}
\gamma_0 &: H(\curl, D)&   \to &  H^{-1/2}(\dive_\Gamma, \Gamma)\\
& u &  \mapsto & u \times \nu
\end{array}
\end{equation*}
is continuous. Moreover, for all $\phi \in H^{-1/2}(\dive_\Gamma, \Gamma)$, there exists $u \in H(\curl, D)$ such that $\gamma_0(u) = \phi$ and 
\begin{equation*}
\| u\|_{H(\curl, D)} \le C \| \phi\|_{H^{-1/2}(\dive_\Gamma, \Gamma)}, 
\end{equation*}
for some positive constant $C$ independent of $\phi$. 
\end{lemma} 

Here 
\begin{equation*}
H^{-1/2}(\dive_\Gamma, \Gamma): = \Big\{ \phi \in [H^{-1/2}(\Gamma)]^3; \; \phi \cdot \nu = 0 \mbox{ and } \dive_\Gamma \phi \in H^{-1/2}(\Gamma) \Big\}
\end{equation*}
\begin{equation*}
\| \phi\|_{H^{-1/2}(\dive_\Gamma, \Gamma)} : = \| \phi\|_{H^{-1/2}(\Gamma)} +  \| \dive_\Gamma \phi\|_{H^{-1/2}(\Gamma)}.
\end{equation*}

Using Lemmas~\ref{lem-compactness} and \ref{lem-trace}, we can easily reach the following result which is used  in the proof of Lemma~\ref{lem-stability} to establish the stability of \eqref{eq-EHDelta}.

\begin{lemma}\label{lem-stability-inside}
 Let $k>0$, $D $ be a smooth open bounded subset of $\mR^3$, $f, g \in [L^2(D)]^3$, and $h_1, h_2 \in H^{-1/2}(\dive_\Gamma, \partial D)$, and let $\eps$ and $\mu$ be two symmetric uniformly elliptic matrix-valued functions defined in $D$ such that \eqref{eps-mu-C1} holds. Assume that  $(\cE, \cH) \in [H( \curl, D)]^2  $ is a  solution to
\begin{equation*}
\left\{ \begin{array}{cll}
\nabla \times \cE = i k  \mu \cH + f& \mbox{ in } D, \\[6pt]
\nabla \times \cH  = - i k \eps \cE + g & \mbox{ in }  D, \\[6pt]
\cH \times \nu = h_1; \; \cE \times \nu = h_2 & \mbox{ on } \partial D.  
\end{array} \right.
\end{equation*}
Then
\begin{equation}\label{stability1-1}
\|(\cE, \cH) \|_{H(\curl, D)} \le C \Big(\|(f, g) \|_{L^2(D)} +  \| (h_1, h_2)\|_{H^{-1/2}(\dive_\Gamma, \partial D)}\Big), 
\end{equation}
for some positive constant $C$ depending on $D$, $\eps$, $\mu$, and $k$ but independent of $f$, $g$, $h_1$, and $h_2$. 
\end{lemma}

\noindent{\bf Proof.} Using Lemma~\ref{lem-trace}, without loss of generality, one may assume that $h_1 = h_2 = 0$. We prove \eqref{stability1-1} by contradiction. Assume that there exist $f_n, g_n \in L^2(D)$ such that 
\begin{equation}\label{contradiction}
\|(\cE^{(n)}, \cH^{(n)}) \|_{H(\curl, D)} = 1 \quad  \mbox{ and } \quad \lim_{n \to + \infty} \|( f_n, g_n)\|_{L^2(D)}  = 0. 
\end{equation}
Here $(\cE^{(n)}, \cH^{(n)})$ is the unique solution to 
\begin{equation}\label{sys-EHn}
\left\{ \begin{array}{cll}
\nabla \times \cE^{(n)} = i k  \mu \cH^{(n)} + f_n& \mbox{ in } D, \\[6pt]
\nabla \times \cH^{(n)}  = - i k \eps \cE^{(n)} + g_n & \mbox{ in }  D, \\[6pt]
\cH^{(n)} \times \nu = \cE^{(n)} \times \nu = 0 & \mbox{ on } \partial D.  
\end{array} \right.
\end{equation}
Applying Lemma~\ref{lem-compactness}, one may assume that $(\cE^{(n)}, \cH^{(n)}) \to (\cE, \cH)$ in $[L^2(D)]^6$ and hence in $[H(\curl, D)]^2$ by \eqref{sys-EHn}. Moreover, 
\begin{equation*}
\left\{ \begin{array}{cll}
\nabla \times \cE = i k  \mu \cH & \mbox{ in } D, \\[6pt]
\nabla \times \cH  = - i k \eps \cE& \mbox{ in }  D, \\[6pt]
\cH \times \nu = \cE  \times \nu = 0 & \mbox{ on } \partial D.  
\end{array} \right.
\end{equation*}
This implies $\cE = \cH = 0$ by the unique continuation principle \cite[Theorem 1]{Tu}. This contradicts the fact 
\begin{equation}
\|(\cE, \cH) \|_{H(\curl, D)} = 1,  
\end{equation}
by \eqref{contradiction}. The conclusion follows. \proofend

\medskip 
We next deal with  the existence,  uniqueness,  and stability of outgoing solutions defined in the whole space. 

\begin{lemma} \label{lem-stability1} Let $k>0$, $D$ be a smooth open bounded subset of $\mR^3$, $f, g \in [L^2(\mR^3)]^3$, $h_1, h_2 \in  H^{-1/2}(\dive_\Gamma, \partial D)$. Assume that $\bar D, \supp f, \supp g \subset B_{R_0}$ for some $R_0> 0$. Let $\eps, \mu$ 
be two symmetric uniformly elliptic matrix-valued functions defined in $\mR^3$ such that \eqref{pro-Identity} and \eqref{eps-mu-C1} hold. There exists  $(\cE, \cH) \in \big[\bigcap_{R>0} H(\curl, B_R \setminus \partial D) \big]^2$ the unique outgoing solution to \footnote{$[ \cdot ]$ denotes the jump across the boundary.}
\begin{equation}\label{haha}
\left\{ \begin{array}{cl}
\nabla \times \cE = i k  \mu \cH + f & \mbox{ in } \mR^3 \setminus \partial D, \\[6pt]
\nabla \times \cH  = - i k \eps \cE + g & \mbox{ in }  \mR^3 \setminus \partial D, \\[6pt]
[\cH \times \nu] = h_1; \; [\cE \times \nu] = h_2 & \mbox{ on } \partial D.  
\end{array} \right.
\end{equation}
Moreover, 
\begin{equation}\label{stability1}
\|(\cE, \cH) \|_{H(\curl, B_{R} \setminus \partial D)} \le C_{R} \Big(\| (f, g) \|_{L^2} + \| (h_1, h_2) \|_{H^{-1/2}(\dive_\Gamma, \partial D)} \Big), 
\end{equation}
for some positive constant $C_R$ depending on $R$, $R_0$, $D$, $\eps$, $\mu$, and $k$,  but independent of $f, g, h_1$, and $h_2$. 
\end{lemma}

The well-posedness of \eqref{haha} is known for $h_1= h_2 = 0$ and $f, g \in H(\dive, \mR^3)$ \footnote{$H(\dive, \mR^3): = \{u \in [L^2(\mR^3)]^3; \; \dive u \in L^2(\mR^3)\}$ and $\| u\|_{H(\dive)} : = \| u \|_{L^2} + \| \dive u\|_{L^2}$} (in this case, $\| (f, g) \|_{L^2} $  is replaced by $\| (f, g) \|_{H(\dive)}$ in \eqref{stability1} since the standard compactness criterion was used). To our knowledge, Lemma~\ref{lem-stability1} is new and the proof requires the new compactness criterion in Lemma~\ref{lem-compactness}.  

\medskip 

\noindent{\bf Proof.} Using Lemma~\ref{lem-trace}, without loss of generality, one may assume that $h_1 = h_2 = 0$. The uniqueness  is a consequence of Rellich's lemma (see,  e.g.,  \cite[Theoren 6.1]{ColtonKressInverse}) and the unique continuation principle \cite[Theorem 1.1]{Tu}. The details are left to the reader. 
The existence and the stability can be derived from the uniqueness  using the limiting absorption principle in the spirit of \cite{Leis} and the compactness result in Lemma~\ref{lem-compactness} as follows. For  $0 < \tau < 1$, let 
$(\cE^\tau, \cH^\tau) \in [H(\curl, \mR^3)]^2$ be the unique  solution to 
\begin{equation}\label{sys-EHfg}
\left\{ \begin{array}{lll}
\nabla \times \cE^\tau &= i k (1 + i \tau) \mu \cH^\tau + f & \mbox{ in } \mR^3, \\[6pt]
\nabla \times \cH^\tau & = - i k (1 + i \tau) \eps \cE^\tau + g & \mbox{ in }  \mR^3. 
\end{array} \right.
\end{equation}
This implies 
\begin{equation*}
\nabla \times \Big( \frac{1}{1 + i \tau}\mu^{-1} \nabla \times \cE^\tau \Big) - k^2 (1 + i \tau) \cE^{\tau}= i k g + \nabla \times \Big( \frac{1}{1 + i \tau}\mu^{-1} f \Big) \mbox{ in } \mR^3. 
\end{equation*}
Multiplying the equation by $\bar \cE^\tau$ (the conjugate of $\cE^{\tau}$), integrating on $\mR^3$, and considering the imaginary part, we have 
\begin{equation*}
\| (\cE^\tau, \cH^\tau)\|_{H(\curl, \mR^3)} \le \frac{C}{\tau}\| (f, g) \|_{L^2}. 
\end{equation*}
Here and in what follows in this proof, $C$ denotes a positive constant independent of $f$, $g$, and $\tau$. 
We claim that 
\begin{equation}\label{claim-Existence}
\| (\cE^\tau, \cH^\tau)\|_{H(\curl, B_{R_0 + 2})} \le C \| (f, g) \|_{L^2}.  
\end{equation}
We prove this by contradiction. Assuming that there exist $\tau_n \to 0_+$ and $f_n, g_n \in L^2(\mR^3)$ with $\supp f_n, \supp g_n \subset B_{R_0}$ such that  
\begin{equation*}
\| (\cE^{(n)}, \cH^{(n)})\|_{H(\curl, B_{R_0 + 2})} = 1 \quad \mbox{ and } \quad \lim_{n \to + \infty}  \| (f_n, g_n)\|_{L^2}= 0. 
\end{equation*}
Here $(\cE^{(n)}, \cH^{(n)}) \in [H(\curl, \mR^3)]^2$ is the unique outgoing solution to \eqref{sys-EHfg} with $f=f_n$, $g= g_n$, and $\tau = \tau_n$. The Stratton-Chu formula (see, e.g., \cite[Theorem 6.6]{ColtonKressInverse}), gives, for $|x| > R_0  + 1$,  
\begin{equation}\label{Chu-1}
\cE^{(n)}(x) = - \curl \int_{\partial B_{R_0 + 1}} (\cE^{(n)} \times \nu) G_n(x, y) \, dy  + \frac{1}{i k_n} \curl \curl \int_{\partial B_{R_0 + 1}} (\cH^{(n)} \times \nu) G_n(x, y) \, dy 
\end{equation}
and
\begin{equation}\label{Chu-2}
\cH^{(n)}(x) = - \curl \int_{\partial B_{R_0 + 1}} (\cH^{(n)} \times \nu) G_n(x, y) \, dy - \frac{1}{i k_n} \curl \curl \int_{\partial B_{R_0 + 1}} (\cE^{(n)} \times \nu) G_n(x, y) \, dy.  
\end{equation}
Here  $k_n= k(1 + i \tau_n)$ and $\dsp G_n (x, y) = \frac{e^{i k_n |x-y|}}{ 4 \pi |x - y|}$.  Since $\| (\cE_n, \cH_n) \|_{L^2(B_{R_0 + 2})}   = 1$, it follows
from \eqref{Chu-1} and \eqref{Chu-2}  that 
\begin{equation*}
\| (\cE^{(n)}, \cH^{(n)})\|_{H(\curl, B_{R})} \le C_{R}, \quad \forall \, R > 0.  
\end{equation*}
Applying Lemma~\ref{lem-compactness}, without loss of generality, one may assume that $(\cE^{(n)}, \cH^{(n)}) \to (\cE, \cH)$ in $[L^2_{\loc}(\mR^3)]^6$, and hence in $[H_{\loc}(\curl, \mR^3)]^2$ by \eqref{sys-EHfg}. Moreover, $(\cE, \cH) \in [H_{\loc}(\curl, \mR^3)]^2$ satisfies 
\begin{equation*}
\left\{ \begin{array}{lll}
\nabla \times \cE &= i k  \mu \cH & \mbox{ in } \mR^3, \\[6pt]
\nabla \times \cH & = - i k \eps \cE & \mbox{ in }  \mR^3. 
\end{array} \right.
\end{equation*}
Letting $n\to \infty$ in \eqref{Chu-1} and \eqref{Chu-2}, we derive that  $(\cE, \cH)$ satisfies the Stratton-Chu formula: 
\begin{equation}\label{Chu-1-1}
\cE(x) = - \curl \int_{\partial B_{R_0 + 1}} (\cE \times \nu) G(x, y) \, dy  + \frac{1}{i k} \curl \curl \int_{\partial B_{R_0 + 1}} (\cH \times \nu) G (x, y) \, dy 
\end{equation}
and
\begin{equation}\label{Chu-2-1}
\cH (x) = - \curl \int_{\partial B_{R_0 + 1}} (\cH \times \nu) G (x, y) \, dy - \frac{1}{i k} \curl \curl \int_{\partial B_{R_0 + 1}} (\cE \times \nu) G(x, y) \, dy, 
\end{equation}
where  $\dsp G (x, y) = \frac{e^{i k |x-y|}}{ 4 \pi |x - y|}$.  Hence $(\cE, \cH)$ satisfies the outgoing condition. The uniqueness of the outgoing solutions yields 
\begin{equation*}
\cE = \cH = 0 \mbox{ in } \mR^3. 
\end{equation*}
This contradicts the fact $ \| (\cE, \cH)\|_{H(\curl, B_{R_0 + 2})} = \lim_{n \to \infty} \| (\cE^{(n)}, \cH^{(n)})\|_{H(\curl, B_{R_0 + 2})} = 1$. 
Hence \eqref{claim-Existence} is proved. From \eqref{claim-Existence}, \eqref{Chu-1} and \eqref{Chu-2}, we obtain 
\begin{equation*}
\| (\cE^\tau, \cH^\tau)\|_{H(\curl, B_{R})} \le C_R \| (f, g) \|_{L^2}.  
\end{equation*}
Applying Lemma~\ref{lem-compactness}, without loss of generality, one may assume that $(\cE^\tau, \cH^\tau) \to (\cE, \cH)$ in $[H_{\loc}(\curl, \mR^3)]^2$ as $\tau \to 0$; moreover, $(\cE, \cH)$ is a  solution to 
\begin{equation*}
\left\{ \begin{array}{lll}
\nabla \times \cE&= i k  \mu \cH + f & \mbox{ in } \mR^3, \\[6pt]
\nabla \times \cH & = - i k  \eps \cE + g & \mbox{ in }  \mR^3. 
\end{array} \right.
\end{equation*}
We also have \eqref{Chu-1-1} and \eqref{Chu-2-1} for $(\cE, \cH)$. Therefore,  $(\cE, \cH)$ satisfies the outgoing condition. 
The estimate of $(\cE, \cH)$ follows from the estimate of $(\cE^\tau, \cH^\tau)$. The proof is complete. \proofend

\begin{remark}   \fontfamily{m} \selectfont The unique continuation of the Maxwell equations has a long story see, e.g., \cite{Protter60, Leis, Tu, BallCapdeboscq} and references therein. It has been known from \cite{Leis}  that the principle holds for $\eps, \mu$ in $C^2$. However, under the assumption $\eps, \mu$ in $C^1$, it was proved  recently in \cite{Tu} (see also \cite{BallCapdeboscq} for a more general setting) using the fact the Maxwell equations can be reduced to a weakly coupled second order elliptic equations see, e.g., \cite[page 168]{Leis}. 
\end{remark}

Similarly, we obtain the following result on the exterior Dirichlet boundary problem.  
\begin{lemma} \label{lem-stability2} Let $k>0$,  $D $ be a smooth open bounded subset of $\mR^3$, $f, g \in [L^2(\mR^3 \setminus D)]^3$, and   $h \in H^{-1/2}(\dive_\Gamma, \partial D)$.  Let $\eps, \mu$ 
be two symmetric uniformly elliptic matrix-valued functions defined in $\mR^3 \setminus D$ such that \eqref{pro-Identity} and \eqref{eps-mu-C1} hold.    Assume that  $\mR^3 \setminus D$ is connected  and $\supp f, \supp g \subset B_{R_0} \setminus D$ for some $R_0>0$. Let   $(\cE, \cH) \in   [H_{\loc}( \curl, \mR^3 \setminus  D)]^2  $ be the unique outgoing solution to 
\begin{equation*}
\left\{ \begin{array}{cll}
\nabla \times \cE = i k  \mu \cH + f & \mbox{ in } \mR^3 \setminus  D, \\[6pt]
\nabla \times \cH  = - i k \eps \cE + g & \mbox{ in }  \mR^3 \setminus D, \\[6pt]
\cE \times \nu = h  & \mbox{ on } \partial D.  
\end{array} \right.
\end{equation*}
Then
\begin{equation*}
\|(\cE, \cH) \|_{H(\curl, B_{R} \setminus D)} \le C_R \Big( \| (f, g) \|_{L^2} +    \| h\|_{H^{-1/2}(\dive_\Gamma, \partial D)} \Big), 
\end{equation*}
for some positive constant $C_R$ depends on $R$, $R_0$, $D$, $\eps$, $\mu$, and $k$, but independent of $f$, $g$, and $h$. 
\end{lemma}

\begin{remark}  \fontfamily{m} \selectfont  The same result holds if the condition $\cE \times \nu = h$ on $\partial D$ is replaced by the condition $\cH \times \nu = h$ on $\partial D$. 
\end{remark}

\noindent{\bf Proof.} The proof of Lemma~\ref{lem-stability2} is similar to the one of Lemma~\ref{lem-stability1}. The details are left to the reader. \proofend

\medskip 
We are ready to state and prove the stability result  for \eqref{eq-EHDelta}.

\begin{lemma} \label{lem-stability} Let $0 < \delta < 1$, $f, g \in [L^2(\mR^3)]^3$, and $(\eps_\delta, \mu_\delta)$ be defined in \eqref{def-eDelta}. Assume that
$\eps$ and $\mu$ are bounded in $\mR^d$ and  satisfy \eqref{pro-eps-mu}, \eqref{pro-Identity} and \eqref{eps-mu-C1}, and $\supp f, \, \supp g,  \,  \bar \Omega_2 \subset B_{R_0}$. 
There exists a unique outgoing solution $({\cE_\delta, \cH_\delta}) \in [H_{\loc}(\curl, \mR^3)]^2$ to 
\begin{equation*}
\left\{ \begin{array}{lll}
\nabla \times \cE_\delta &= i k  \mu_\delta \cH_\delta + f & \mbox{ in } \mR^3,\\[6pt]
\nabla \times \cH_\delta & = - i k  \eps_\delta \cE_\delta + g & \mbox{ in }  \mR^3. 
\end{array} \right.
\end{equation*}
Moreover,  
\begin{equation}\label{stability1-11}
\|(\cE_\delta, \cH_\delta) \|_{H(\curl, B_{R})} \le \frac{C_{R}}{\delta} \|( f, g) \|_{L^2}. 
\end{equation}
Assume in addition that $\supp f \subset \bar D$, $\supp g \subset \bar D$, and $ \bar D \cap \Omega_2 = \O$ for some smooth open subset $D$ of $\mR^3$. Then 
\begin{equation}\label{stability2}
\|(\cE_\delta, \cH_\delta) \|_{H(\curl, B_{R})}^2 \le \frac{C_{R}}{\delta} \| (f, g)\|_{L^2} \|(\cE_\delta, \cH_\delta) \|_{H(\curl, D)} + C_R \|( f, g) \|_{L^2}^2, 
\end{equation}
Here $C_R$ denotes a positive constant depending on $R$, $R_0$, $\eps$, $\mu$, and $D$ but independent of $f$, $g$, and $\delta$. 
\end{lemma}

\begin{remark}  \fontfamily{m} \selectfont 
 Lemma~\ref{lem-stability} does not require  any assumptions on the reflecting complementary property. In the proof of Theorem~\ref{thm1}, we apply Lemma~\ref{lem-stability} with $D =B_R \setminus \Omega_2$ for some $R> 0$. 
\end{remark}

\noindent{\bf Proof.} For $\delta > 0$ fixed, the existence and uniqueness of $(\cE_\delta, \cH_\delta)$ can be obtained as in the proof of Lemma~\ref{lem-stability1}. The details are omitted.  We only give the proof of \eqref{stability1-11} and \eqref{stability2}. We have, in $\mR^3$,  
\begin{equation*}
\nabla \times ( \mu_\delta^{-1} \nabla \times \cE_\delta) - k^2 \eps_\delta \cE_\delta =  \nabla \times (\mu_\delta^{-1} f) +  i k g. 
\end{equation*}
Set 
\begin{equation*}
M_\delta = \frac{1}{\delta} \|(f, g) \|_{L^2} \| (\cE_\delta, \cH_\delta) \|_{L^2(B_{R_0})} + \|(f, g)\|_{L^2}^2. 
\end{equation*}
Multiplying the equation by $\bar \cE_\delta$, integrating in $B_R$, and using the fact $\supp f \subset B_{R_0}$, we have, for $R > R_0$, 
\begin{multline*}
\int_{B_R} \langle  \mu_\delta^{-1} \nabla \times \cE_\delta, \nabla \times \cE_\delta \rangle  - \int_{\partial B_{R}}  \langle (\mu_\delta^{-1} \nabla \times \cE_\delta) \times \nu, \cE_\delta \rangle  - k^2 \int_{B_R} \langle \eps_\delta \cE_\delta, \cE_\delta \rangle  \\[6pt]
=  \int_{B_R}  \langle \mu_\delta^{-1} f, \nabla \times \cE_\delta \rangle  +  \int_{B_R}  \langle i k g, \cE_\delta \rangle.
\end{multline*}
Since  $\mu_{\delta} = I$, $f = 0$, and $\nabla \times  \cE_\delta = i k \cH_\delta $ in $\mR^3 \setminus B_{R_0}$, we derive that, for $R> R_0$,  
\begin{multline*}
\int_{B_R} \langle  \mu_\delta^{-1} \nabla \times \cE_\delta, \nabla \times \cE_\delta \rangle  + \int_{\partial B_{R}}  \langle i k H_\delta , \cE_\delta \times \nu \rangle  - k^2 \int_{B_R} \langle \eps_\delta \cE_\delta, \cE_\delta \rangle  \\[6pt]
=  \int_{B_R}  \langle \mu_\delta^{-1} f, \nabla \times \cE_\delta \rangle  +  \int_{B_R}  \langle i k g, \cE_\delta \rangle.
\end{multline*}
Letting $R \to + \infty$, using the outgoing condition,  and considering the imaginary part,  we obtain 
\begin{equation}\label{inE}
 \|\cE_\delta \|_{H(\curl, \Omega_2 \setminus \Omega_1)}^2 \le C M_\delta.    
\end{equation}
This  implies, by Lemma~\ref{lem-trace},  
\begin{equation}\label{bdryE}
\| \cE_\delta \times \nu \|_{H^{-1/2}(\dive_\Gamma, \partial \Omega_2)}^2 + \| \cE_\delta \times \nu \|_{H^{-1/2}(\dive_\Gamma, \partial \Omega_1)}^2  \le C M_\delta. 
\end{equation}
Using the equations of $(\cE_\delta, \cH_\delta)$, we derive from \eqref{inE} that 
\begin{equation}\label{inH}
 \|\cH_\delta \|_{H(\curl, \Omega_2 \setminus \Omega_1)}^2 \le C M_\delta; 
\end{equation}
which yields, by Lemma~\ref{lem-trace} again,  
\begin{equation}\label{bdryH}
\| \cH_\delta \times \nu \|_{H^{-1/2}(\dive_\Gamma, \partial \Omega_2)}^2 + \| \cH_\delta \times \nu \|_{H^{-1/2}(\dive_\Gamma, \partial \Omega_1)}^2 \le C M_\delta. 
\end{equation}
Applying Lemma~\ref{lem-stability2}, we have 
\begin{equation}\label{inEHo}
\| (\cE_\delta, \cH_\delta)\|_{H(\curl, B_R \setminus \Omega_2)}^2 \le C_R M_\delta,
\end{equation}
and applying Lemma~\ref{lem-stability-inside}, we obtain
\begin{equation}\label{inEHi}
\| (\cE_\delta, \cH_\delta)\|_{H(\curl,  \Omega_1)}^2 \le C M_\delta. 
\end{equation}
A combination of \eqref{inE}, \eqref{inH}, \eqref{inEHo}, and \eqref{inEHi} yields 
\begin{equation}\label{hahahaha}
\| (\cE_\delta, \cH_\delta)\|_{H(\curl, B_R)} \le C_R M_\delta. 
\end{equation}
This implies  \eqref{stability1-11}. Inequality \eqref{stability2} follows from  \eqref{hahahaha} by noting that in the definition of  $M_\delta$, one can replace $\|(E_\delta, H_\delta) \|_{L^2}$ by $\|(E_\delta, H_\delta)\|_{L^2(D)}$ if $\supp f \subset \bar D$, $\supp g \subset \bar D$, and $ \bar D \cap \Omega_2 = \O$.    \proofend

\medskip 
The following change of variables for the Maxwell equations motivates the definition of reflecting complementary media. 

\begin{lemma} \label{lem-CV}Let $D, D'$ be two open bounded connected subsets of $\mR^3$ and ${\cal T}: D \to D'$ be  bijective such that 
${\cal T} \in C^1(\bar D)$ and ${\cal T}^{-1} \in C^1(\bar D')$. Assume that $\eps, \mu \in [L^\infty(D)]^{3 \times 3}$,  $j \in [L^2(D)]^3$ and $(E, H) \in [H(\curl, D)]^2$ is a solution to 
\begin{equation*}\left\{
\begin{array}{lll}
\nabla \times E  &=  i k \mu H & \mbox{ in } D, \\[6pt]
\nabla \times H &=  - i k \eps  E + j & \mbox{ in } D. 
\end{array}\right. 
\end{equation*}
Define $(E', H')$ in $D'$ as follows 
\begin{equation}\label{TEH}
E'(x') = {\cal T}*E(x'): = \nabla {\cal T}^{-T}(x) E(x) \mbox{  and  } H'(x') = {\cal T}*H(x'): = \nabla {\cal T}^{-T}(x) H(x),  
\end{equation}
with $x' = {\cal T}(x)$ and set $\eps' = {\cal T}_*\eps$, $\mu' = {\cal T}_*\mu$, and $j'= T_*j$ by \eqref{def-TT}.  Then $(E', H')$ is a solution to 
\begin{equation}\label{interior-Lem}
\left\{
\begin{array}{lll}
\nabla' \times E'  &=  i k \mu' H' & \mbox{ in } D', \\[6pt] 
\nabla' \times H' &=  - i k \eps'  E' + j' & \mbox{ in } D', 
\end{array}\right. 
\end{equation}
Assume in addition that  $D$ is of class $C^1$ and ${\bf T} = {\cal T} \big|_{\partial D}: \partial D \to \partial D'$ is a diffeomorphism. 
Let  $\nu$ and $\nu'$ denote the outward unit normal  vector on $\partial D$ and $\partial D'$.
We have 
\begin{equation}\label{bdry}
\mbox{if } E \times \nu = g  \mbox{ and } H \times \nu = h \mbox{ on } \partial D \mbox{ then } E' \times \nu' = {\bf T}_*g \mbox{ and } H' \times \nu' = {\bf T}_* h  \mbox{ on } \partial D', 
\end{equation}
where ${\bf T}_*$ is defined by, for a tangential vector field $\varphi$  defined in $\partial D$,   
\begin{equation}\label{def-T}
{\bf T}_*\varphi(x') = \sign \cdot \frac{\nabla_{\partial D} {\bf T}(x) \varphi(x)}{|\det \nabla_{\partial D}  {\bf T}(x)|}  \mbox{ with } x' = {\bf T}(x),  
\end{equation}
where $\sign := \det \nabla {\cal T} (x)/ |\det \nabla {\cal T}(x)|$ for some $x \in D$. In particular, if $D \cap D' = \O$ and ${\cal T}(x) = x$ on $\partial D \cap \partial D'$, then 
\begin{equation}\label{reflection}
H' \times \nu = H \times \nu \quad \mbox{ and } \quad E' \times \nu = E \times \nu  \quad \mbox{ on } \partial D \cap \partial D'. 
\end{equation}
\end{lemma}


Assertion \eqref{reflection} immediately  follows from \eqref{bdry}   by noting that  $\sign = -1$, $\nu' = - \nu$, and ${\bf T} = I$ on $\partial D \cap \partial D'$ in this case. This assertion is  used several times in the proof of Theorem~\ref{thm1}.

\begin{remark} \fontfamily{m} \selectfont
Note that, the definition of ${\cal T}*$ is different from ${\cal T}_*$ for a field in $\mR^3$. It is helpful to remember that for electromagnetic fields  \eqref{TEH} is used whereas for sources \eqref{def-TT} is involved.   
\end{remark}

\begin{remark}  \fontfamily{m} \selectfont  System \eqref{interior-Lem} is known for a smooth pair $(E, H)$.   
Statement \eqref{bdry}  might be known; however, we cannot find a reference for it.  For the convenience of the reader, we give  the details of the proof in Appendix A for the  form stated here. 
\end{remark}

\subsection{Proof of Theorem~\ref{thm1}} \label{sect-complementary}

The proof  is divided into three steps. 
\begin{itemize}
\item Step 1: Assume that there exists an outgoing solution $(E_0, H_0) \in [H_{\loc}(\curl, \mR^3)]^2$  of \eqref{limit-EH-0}.
Then $j$ is compatible and $(E_0, H_0) = (\crE_0, \crH_0)$, where 
\begin{equation}\label{def-NI} (\crE_0, \crH_0) := \left\{
\begin{array}{cl}
(\hE, \hH) & \mbox{ in } \mR^3 \setminus \Omega_3, \\[6pt]
(\bE, \bH) &\mbox{ in }  \Omega_3 \setminus \Omega_2, \\[6pt]
 (F^{-1}* \bE, F^{-1}* \bH) & \mbox{ in } \Omega_2 \setminus \Omega_1, \\[6pt]
(F^{-1}*G^{-1}* \hE, F^{-1}* G^{-1}*\hH) & \mbox{ in } \Omega_1. 
\end{array}\right.
\end{equation}

\item Step 2: Assume that $j$ is compatible. Then $(\crE_0, \crH_0)$ given in \eqref{def-NI} is the unique outgoing solution of \eqref{limit-EH-0}. Moreover, 
$$
\| (E_\delta, H_\delta) - (\crE_0, \crH_0) \|_{H(\curl, B_R)} \le C_R \delta^{1/2} \| (\crE_0, \crH_0)\|_{L^2\big( (\Omega_2 \setminus \Omega_1) \cup B_R\big)}. 
$$

\item Step 3:  Assume that $j$ is not compatible. Then 
\begin{equation*}
\lim_{\delta \to 0_+} \| (E_\delta, H_\delta)\|_{H(\curl, B_R)}  = + \infty, 
\end{equation*}
for $R>0$ such that  $\bar \Omega_2 \subset B_R$. 
\end{itemize}

It is clear that the conclusion follows after Step 3.  We now proceed these steps. 
\medskip

\noindent \underline{Step 1:}  Let  $(\oE_0, \oH_0)$  be the reflection of $(E_0, H_0)$ in $\Omega_2$ through $\partial \Omega_2$ by F, i.e., \footnote{The definition of $F*E$ and $F*H$ are given in \eqref{TEH}.}
\begin{equation*}
(\oE_0,  \oH_0 )  = (F*E_0, F* H_0) \mbox{ in } \mR^3 \setminus \Omega_2; 
\end{equation*}
which implies 
\begin{equation}\label{part1-Step1}
(E_0,  H_0 )  = (F^{-1}*\oE_0, F^{-1}* \oH_0) \mbox{ in } \Omega_2 \setminus \{x_1\}. 
\end{equation}
Recall  that $(\eps_0, \mu_0) = (\eps, \mu)$ in $\mR^3$.  It follows from Lemma~\ref{lem-CV} that 
\begin{equation}\label{E1H1}
\left\{ \begin{array}{llll}
\nabla \times \oE_0 &=&  i k  F_* \mu \oH_0 & \mbox{ in } \mR^3 \setminus \Omega_2, \\[6pt]
\nabla \times \oH_0 & = & - i k  F_*\eps \oE_0 & \mbox{ in }  \mR^3 \setminus \Omega_2, 
\end{array} \right.
\end{equation}
and 
\begin{equation*}
\oE_0 \times \nu  = E_0 \times \nu \quad \mbox{ and } \quad \oH_0 \times \nu  = H_0 \times \nu  \quad \mbox{ on } \partial \Omega_2. 
\end{equation*}
Since $(F_*\eps, F_*\mu) = (\eps, \mu)$ in $\Omega_3 \setminus \Omega_2$, it follows  from the unique continuation principle that 
\begin{equation}\label{E0E1}
(\oE_0, \oH_0) = (E_0, H_0) \mbox{ in } \Omega_3 \setminus \Omega_2.
\end{equation}
Let $(\tE_0, \tH_0)$  be the reflection of $(\oE_0, \oH_0)$ in $\mR^3 \setminus \Omega_3$ through $\partial \Omega_3$ by G, i.e., 
\begin{equation*}
(\tE_0,  \tH_0)  = (G*\oE_0,   G*\oH_0) \mbox{ in } \Omega_3 \setminus \{x_1\}; 
\end{equation*}
which implies 
\begin{equation}\label{part2-Step1}
(\oE_0,  \oH_0 )  = (G^{-1}*\tE_0, G^{-1}* \tH_0) \mbox{ in } \mR^3 \setminus \bar \Omega_3. 
\end{equation}
Set 
\begin{equation*} 
(\cE, \cH) = \left\{\begin{array}{cl} (E_0, H_0) & \mbox{ in } \mR^3 \setminus \Omega_3, \\[6pt] 
(\tE_0, \tH_0) & \mbox{ in } \Omega_3. 
\end{array}\right. 
\end{equation*}
We have, by applying  Lemma~\ref{lem-CV} and using \eqref{E1H1}, 
\begin{equation*}
\left\{ \begin{array}{llll}
\nabla \times \tE_0 &=&  i k  \hat \mu \tH_0 & \mbox{ in }  \Omega_3, \\[6pt]
\nabla \times \tH_0 & = & - i k  \hat  \eps \tE_0 & \mbox{ in }   \Omega_3, 
\end{array} \right.
\end{equation*}
and, by applying  Lemma~\ref{lem-CV} and using \eqref{E0E1}, 
\begin{equation*}
\tE_0 \times \nu  = \oE_0 \times \nu  = E_0 \times \nu  \quad \mbox{ and } \quad  \tH_0 \times \nu  =  \oH_0 \times \nu  =  H_0 \times \nu   \quad \mbox{ on } \partial \Omega_3. 
\end{equation*}
It follows that  $(\cE, \cH) \in [H_{\loc}(\curl, \mR^3)]^2$, $(\cE, \cH)$ satisfies the outgoing condition,  and  
\begin{equation*}
\left\{ \begin{array}{clll}
\nabla \times \cE =  i k \hat  \mu \cH & \mbox{ in }  \mR^3, \\[6pt]
\nabla \times \cH  =  - i k \hat \eps \cE +  j & \mbox{ in }  \mR^3, 
\end{array} \right.
\end{equation*}
We derive that 
\begin{equation*}
(\cE, \cH) = (\hE, \hH) \mbox{ in } \mR^3 \quad \mbox{ and } \quad  
(\bE, \bH) = (E_0, H_0) \mbox{ in } \Omega_3 \setminus \Omega_2. 
\end{equation*}
From the definitions of $(\crE_0, \crH_0)$ and $(\cE, \cH)$, \eqref{part1-Step1}, and \eqref{part2-Step1}, we obtain  
\begin{equation*} (E_0, H_0) = (\crE_0, \crH_0) \mbox{ in } \mR^3.  
\end{equation*}

\medskip 
\noindent \underline{Step 2:}  It is clear that 
\begin{equation}\label{state1-Step2}
(\crE_0, \crH_0) \in [H(\curl, B_R \setminus (\partial \Omega_1 \cup \partial \Omega_2 \cup \partial \Omega_3)) \mbox{ for all } R>0. 
\end{equation}
Using the fact $F(x) =x$ on $\partial \Omega_2$ and $G(x) = x$ on $\partial \Omega_3$ and applying Lemma~\ref{lem-CV}, we have 
\begin{equation}\label{state2-Step2}
[\crE_0 \times \nu] = [\crH_0 \times \nu] = 0 \mbox{ on } \partial \Omega_1 \cap \partial \Omega_2.
\end{equation}
From the definition of $({\bE, \bH})$ and $(\crE, \crH)$, we obtain 
\begin{equation}\label{state3-Step2}
[\crE_0 \times \nu] = [\crH_0 \times \nu] = 0 \mbox{ on }  \partial \Omega_3.
\end{equation}
Applying Lemma~\ref{lem-CV} again, we get
\begin{equation} \label{state4-Step2}
\left\{ \begin{array}{clll}
\nabla \times \crE_0 =  i k \hat  \mu \crH_0 & \mbox{ in }  \mR^3 \setminus  (\partial \Omega_1 \cup \partial \Omega_2 \cup \partial \Omega_3), \\[6pt]
\nabla \times \crH_0  =  - i k \hat \eps \crE_0 +  j & \mbox{ in }  \mR^3 \setminus (\partial \Omega_1 \cup \partial \Omega_2 \cup \partial \Omega_3). 
\end{array} \right.
\end{equation}
We derive from \eqref{state1-Step2}, \eqref{state2-Step2}, \eqref{state3-Step2}, and \eqref{state4-Step2} that 
$(\crE_0, \crH_0) \in [H_{\loc}(\curl, \mR^3)]^2$ is an outgoing solution of \eqref{limit-EH-0} and hence the unique outgoing solution by Step 1.  We have
\begin{equation*}
\left\{ \begin{array}{lll}
\nabla \times (E_\delta - \crE_0) &= i k  \mu_\delta (H_\delta - \crH_0) + i k (\mu_0 - \mu_\delta) \crH_0& \mbox{ in } \mR^3, \\[6pt]
\nabla \times (H_\delta - \crH_0) & = - i k  \eps_\delta (E_\delta - \crE_0) + i k (\eps_\delta- \eps_0)  \crE_0 & \mbox{ in }  \mR^3. 
\end{array} \right.
\end{equation*}
Moreover, $(E_\delta - \crE_0, H_\delta -\crH_0)$ satisfies the outgoing condition. 
Applying \eqref{stability1-11} in Lemma~\ref{lem-stability}, we have, for $R> 0$,  
\begin{equation*}
\| (E_\delta - \crE_0, H_\delta - \crH_0)\|_{H(\curl, B_R)} \le C_R \|(\crE_0, \crH_0) \|_{L^2(\Omega_2 \setminus \Omega_1)}; 
\end{equation*}
which implies 
\begin{equation}\label{state5-Step2}
\| (E_\delta, H_\delta)\|_{H(\curl, B_R)} \le C_R \|(\crE_0, \crH_0) \|_{L^2\big((\Omega_2 \setminus \Omega_1) \cup B_R\big)}. 
\end{equation}
Applying \eqref{stability2} in Lemma~\ref{lem-stability} for $(E_\delta - \crE_0, H_\delta - \crH_0)$ and using \eqref{state5-Step2}, we obtain this time, for $R> 0$,  
\begin{equation*}
\| (E_\delta - \crE_0, H_\delta - \crH_0)\|_{H(\curl, B_R)} \le C_R \delta^{1/2} \|(E_0, H_0) \|_{L^2\big((\Omega_2 \setminus \Omega_1) \cup B_R \big)}. 
\end{equation*}

\medskip 
\noindent \underline{Step 3:} We prove Step 3 by contradiction. Assume that there exists $R>0$ such that $\bar \Omega_2 \subset B_R$,  and,  for some $(\delta_n) \to 0$, 
\begin{equation*}
\sup_{n} \| (E_{\delta_n}, H_{\delta_n})\|_{H(\curl, B_R)}  < + \infty.  
\end{equation*}
Applying Lemma~\ref{lem-stability2}, we have 
\begin{equation*}
\sup_{n} \| (E_{\delta_n}, H_{\delta_n})\|_{H(\curl, B_r)}  < + \infty \mbox{ for all } r > 0.   
\end{equation*}
Without loss of generality, one may assume that $(E_{\delta_n}, H_{\delta_n}) \wc (E_0, H_0)$ weakly in \\$[H_{\loc}(\curl, \mR^3)]^2$. Then $(E_0, H_0)$ is an outgoing solution to \eqref{limit-EH-0}. Therefore, $j$ is compatible by Step 1. We have a contradiction. The proof of Step 3 is complete. 
\proofend

\begin{remark} \fontfamily{m} \selectfont  It is clear from the proof that in the case $j$ is compatible,  $(E_0, H_0)$ can be determined by: 
\begin{equation}\label{def-NI-1} (E_0, H_0) = \left\{
\begin{array}{cl}
(\hE, \hH) & \mbox{ in } \mR^3 \setminus \Omega_3, \\[6pt]
(\bE, \bH) &\mbox{ in }  \Omega_3 \setminus \Omega_2, \\[6pt]
 (F^{-1}* \bE, F^{-1}* \bH) & \mbox{ in } \Omega_2 \setminus \Omega_1, \\[6pt]
(F^{-1}*G^{-1}* \hE, F^{-1}* G^{-1}*\hH) & \mbox{ in } \Omega_1. 
\end{array}\right.
\end{equation}
This is the key observation for the setting of Theorem~\ref{thm1}. 
\end{remark}

\begin{remark} \fontfamily{m} \selectfont In this paper, we confine ourself to the case $\supp j \cap \Omega_3 = \O$ for simple presentation. In fact,  the proof of Theorem~\ref{thm1}  can be extended to cover the  case  where no condition on $\supp j$ is required. The details are left to the reader (see \cite{Ng-Complementary} for a complete account in the acoustic setting). 
\end{remark}

\begin{remark}  \fontfamily{m} \selectfont In \cite[Theorems 2 and 3 and  Proposition 2]{Ng-WP} we showed that in the acoustic setting the complementary property is ``necessary" to the appearance of  resonance in the sense that the field can blow up  in $L^2$-norm even in the region away from the interface of sign changing coefficients. This property would hold in the electromagnetic setting
and will be considered elsewhere. 
\end{remark}

\section{Proof of Theorem~\ref{thm-lens}} \label{sect-thm-lens}

Theorem~\ref{thm-lens} is a consequence of Theorem~\ref{thm1}.  In fact, we can derive from Theorem~\ref{thm1} the  following more general result: 

\begin{proposition}\label{pro-DCM} Let $0 < \delta < 1$,  $j \in L^2_{\mc}(\mR^3)$  and let $(E_\delta, H_\delta) \in [H_{\loc}(\curl, \mR^3)]^2$ be the unique outgoing solution of \eqref{eq-EHDelta}. 
Assume that  $(\eps, \mu)$ in $\Omega_{2} \setminus \Omega_{1}$ and $(\eps,  \mu)$ in $\Omega_{3} \setminus \Omega_{2}$  are  reflecting complementary for some $\Omega_2 \subset \subset \Omega_3 \subset \subset \mR^3$,  {\bf and} $(\hat \eps, \hat \mu) = (\eps, \mu)$ in $\Omega_3 \setminus \Omega_2$.  Then $j$ with  $\supp j \cap \Omega_3 = \O$ is compatible and  there exists a unique outgoing solution  $(E_0, H_0) \in [H_{\loc}(\curl, \mR^3)]^2$ to
\begin{equation}\label{limit-EH-0}
\left\{ \begin{array}{lll}
\nabla \times E_0 &= i k  \mu H_0 & \mbox{ in } \mR^3, \\[6pt]
\nabla \times H_0 & = - i k  \eps E_0 + j & \mbox{ in }  \mR^3. 
\end{array} \right.
\end{equation}
Moreover, 
$$
(E_0, H_0) = (\hE, \hH) \mbox{ in } \mR^3 \setminus \Omega_3, 
$$
and, for all $R>0$, 
$$
\| (E_\delta, H_\delta) - (E_0, H_0) \|_{H(\curl, B_R)} \le C_R \delta^{1/2 } \| j\|_{L^2}, 
$$
for some positive constant $C_R$ independent of $j$ and $\delta$. 
\end{proposition}

\noindent{\bf Proof.} Since $(\hat \eps, \hat \mu) = (\eps, \mu)$ in $\Omega_3 \setminus \Omega_2$, we have 
\begin{equation*}
\left\{\begin{array}{cllll}
\nabla \times \hE = i k  \mu \hH & \mbox{ in } \Omega_3 \setminus \Omega_2,  \\[6pt]
\nabla \times \hH  = - i k  \eps \hE  & \mbox{ in } \Omega_3 \setminus \Omega_2.   
\end{array}\right.
\end{equation*}
Hence $(\bE, \bH)$ exists in $\Omega_3 \setminus \Omega_2$ and $(\bE , \bH) = (\hE, \hH)$ in $\Omega_3 \setminus \Omega_2$.  We derive from \eqref{def-NI-1}
that 
 \begin{equation}\label{def-NI-2} (E_0, H_0) = \left\{
\begin{array}{cl}
(\hE, \hH) & \mbox{ in } \mR^3 \setminus \Omega_2, \\[6pt]
 (F^{-1}* \hE, F^{-1}* \hH) & \mbox{ in } \Omega_2 \setminus \Omega_1, \\[6pt]
(F^{-1}*G^{-1}* \hE, F^{-1}* G^{-1}*\hH) & \mbox{ in } \Omega_1. 
\end{array}\right.
\end{equation}
On the other hand, from the definition of $(\hE, \hH)$ and Lemma~\ref{lem-stability1}, we have 
$$
\| (\hE, \hH) \|_{H(\curl, B_R)} \le C_R \| j\|_{L^2}. 
$$
It follows from \eqref{def-NI-2} that 
$$
\| (E_0, H_0) \|_{H(\curl, B_R)} \le C_R \| j\|_{L^2}. 
$$
The conclusion now follows from Theorem~\ref{thm1}. \proofend

\medskip 
We are now ready to give 
\medskip 

\noindent{\bf Proof of Theorem~\ref{thm-lens}.} Applying  Proposition~\ref{pro-DCM} with $\Omega_j = B_{r_j}$ and noting that $(\hat \eps, \hat  \mu) = (\eps, \mu)$ in $B_{r_3} \setminus B_{r_2}$ by \eqref{the-choice}, one obtains the conclusion of Theorem~\ref{thm-lens}. \proofend

\begin{remark} \fontfamily{m} \selectfont 
It follows from \eqref{choice-123} that $r_3 \to m r_0$
as $\alpha \to 1_+$. Therefore, for any $\eps > 0$, there exists a lens-construction such that $m$ times  magnification for an  object in $B_{r_0}$  takes place for any $f$ with $\supp f \cap  B_{mr_0 + \eps} = \O$. 
\end{remark}

\begin{remark} \fontfamily{m} \selectfont
The construction of lenses is not restricted to the symmetric geometry considered here: the geometry of lenses can be quite arbitrary (see Proposition~\ref{pro-DCM}). This is one of the motivations of the study reflecting complementary media in a general form given in Section~\ref{sect-2}. 
\end{remark}

\begin{remark} \fontfamily{m} \selectfont In comparison with the lens construction in \cite{Ng-Superlensing} for the acoustic setting, we assume here that $m r_0 = r_2$ instead of the condition $mr_0 < \sqrt{r_2 r_3}$. A rate of the convergence which is $\delta^{1/2}$  is obtained in this case and the proof does not involve the removing localized singularity technique. 
\end{remark}

In the rest of this section, we give 

\medskip 
\noindent{\bf Proof of \eqref{formula-eps-mu}.} We have
\begin{equation*}
F^{-1}(x) = \frac{r_3^\beta}{m} K(x) \mbox{ where } K(x) =  \frac{x}{|x|^\beta}. 
\end{equation*}
We claim that 
\begin{equation}\label{computation-part1}
\frac{\nabla K(x) \nabla K(x)^T}{\det (\nabla K(x))} =- |x|^\beta \Big[ \frac{1}{\alpha -1} e_r \otimes e_r + (\alpha-1) \big(e_\theta \otimes e_\theta + e_\varphi \otimes e_\varphi \big)\Big]. 
\end{equation}
By using rotations, it suffices to prove \eqref{computation-part1} for  $x =(x_1, 0, 0)$. We have 
\begin{equation*}
\frac{\partial}{\partial x_j} \Big( \frac{x_i}{|x|^\beta}\Big)  = \frac{\delta_{ij}}{|x|^{\beta}} - \beta \frac{x_i x_j}{|x|^{\beta + 2}}. 
\end{equation*} 
 It follows that, for $x= (x_1, 0, 0)$,  
\begin{align*}
\frac{\nabla K(x) \nabla K(x)^T}{\det (\nabla K(x))} & = \frac{|x_1|^\beta}{(1 - \beta)}\left(\begin{array}{ccc} (1 - \beta)^2 & 0 & 0 \\[6pt]
0 & 1 & 0 \\[6pt]
0 & 0 & 1
\end{array}\right); 
\end{align*}
hence \eqref{computation-part1} is proved since $(\alpha -1)(\beta -1) = 1$. 
Thus,  for $x'=F^{-1}(x)$, we have
\begin{align*}
F^{-1}_*I(x') = & - \frac{m}{r_3^\beta} \frac{r_2^{\alpha \beta }}{|x'|^{\beta (\alpha -1)}} \Big[ \frac{1}{\alpha -1} e_{r'} \otimes e_{r'} + (\alpha-1) \big(e_{\theta'} \otimes e_{\theta'} + e_{\varphi'} \otimes e_{\varphi'} \big)\Big], 
\end{align*}
since $(e_{r'}, e_{\theta'}, e_{\varphi'}) = (e_r, e_{\theta}, e_{\varphi})$. 
Using the fact that $\beta (\alpha -1 ) = \alpha$ and 
\begin{equation*}
\frac{m r_2^{\alpha \beta}}{r_3^\beta} = r_2^\alpha \frac{m r_2^{\alpha(\beta -1)}}{r_3^\beta} = r_2^\alpha \frac{m r_2^{\beta}}{r_3^\beta} = r_2^\alpha m \left( \frac{1}{m^\frac{\alpha -1}{\alpha}}\right)^\beta = r_2^\alpha, 
\end{equation*}
we obtain
\begin{equation*}  
F^{-1}_*I(x') =  - \frac{r_2^\alpha}{|x'|^{\alpha}} \Big[ \frac{1}{\alpha -1} e_{r'} \otimes e_{r'} + (\alpha-1) \big(e_{\theta'} \otimes e_{\theta'} + e_{\varphi'} \otimes e_{\varphi'} \big)\Big]; 
\end{equation*}
which is \eqref{formula-eps-mu}. \proofend


\appendix
\section{Appendix: Proof of Lemma~\ref{lem-CV}}
\renewcommand{\theequation}{A\arabic{equation}}
\renewcommand{\thelemma}{A\arabic{lemma}}
  \setcounter{lemma}{0}  

Set 
\begin{equation*}
J(x) = \nabla T(x) \mbox{ for } x \in D. 
\end{equation*}
We first prove $\nabla' \times H' =  - i k \eps'  E' + j' $ for smooth $(E, H)$. 
We have, for indices in $\{1, 2, 3\}$,  
\begin{equation*}
(\nabla \times H)_c =  \epsilon_{abc} \partial_a H_b, \quad \partial_a = J_{da} \partial_d',  
\end{equation*}
and
\begin{equation*}
 \quad H_b = (J^T H')_b = J_{db} H'_d, \quad  E_c = (J^T E')_c = J_{dc} E'_d.  
\end{equation*}
Here $\epsilon_{abc}$ denotes the usual Levi Civita permutation, i.e., 
\begin{equation*}
\epsilon_{abc} = \left\{\begin{array}{cl} \sign(abc) & \mbox{ if } abc \mbox{ is a permutation}, \\[6pt]
0 & \mbox{ otherwise.}
\end{array} \right.
\end{equation*}
Then 
\begin{equation}\label{p1-1}
(\nabla \times H)_c = \epsilon_{abc} \partial_a H_b = \epsilon_{abc} J_{da} \partial_d' (J_{eb} H'_e) = \epsilon_{abc} J_{da} J_{eb} \partial_d' H'_e
\end{equation}
and 
\begin{equation}\label{p2-1}
- i k (\eps E)_c = - i k \eps_{c d} E_d = - i k \eps_{cd } J_{ed} E'_e. 
\end{equation}
Here in the last identity of \eqref{p1-1}, we used the fact that 
\begin{equation*}
\epsilon_{abc} J_{da} \partial_d' (J_{eb}) = 0. 
\end{equation*} 
From \eqref{p1-1}, we derive that 
\begin{equation*}
J_{fc}(\nabla \times H)_c  = \epsilon_{abc}  J_{fc }J_{da} J_{eb} \partial_d' H'_e = \det J \,  \epsilon_{def}   \partial_d' H'_e = \det J ( \nabla' \times H')_f
\end{equation*}
which yields 
\begin{equation}\label{p1-2}
J (\nabla \times H) = \det J (\nabla' \times H'). 
\end{equation}
From \eqref{p2-1}, we obtain 
\begin{equation*}
- i k J_{fc} (\eps E)_c = - i k J_{fc} \eps_{cd } J_{ed} E'_e = - ik \det J (\eps' E')_f; 
\end{equation*}
which implies 
\begin{equation}\label{p2-2}
- i k J \eps E = - ik \det J \eps' E'.  
\end{equation}
It is clear that 
\begin{equation}\label{p3-1}
J_{fc} j_c = \det J j'_f. 
\end{equation}
Identity $\nabla' \times H' = - i k \eps' E'  + j'$ for smooth $(E, H, j)$  now follows from \eqref{p1-2}, \eqref{p2-2}, and \eqref{p3-1}. The proof of $\nabla' \times H' =  - i k \eps'  E' + j' $ in the general case $(E, H) \in [H(\curl, D)]^2$ can be proceeded   as follows. We have, for $\varphi \in [C^1_{\mc}(D)]^3$,  
\begin{equation}\label{p1}
\int_{D} (\nabla \times H) J^T \varphi = \int_{D} (- i k \eps E + j) J^T \varphi. 
\end{equation}
It is clear  that 
\begin{equation}\label{p2}
\int_{D} (- i k \eps E + j) J^T \varphi = \int_{D} (- i k J \eps J^T J^{-T} E + Jj) \varphi = \sign \cdot  \int_{D'} (- i k \eps' E' + j') \varphi_1,  
\end{equation}
where $\varphi_1(x') = \varphi(x)$. In the last inequality, we made a change of variable $x'  = T(x)$. On the other hand, 
\begin{equation*}
\int_{D} (\nabla \times H) J^T \varphi  = -  \int_{D}  H \nabla \times ( J^T \varphi). 
\end{equation*}
We have, as in \eqref{p1-2},  
\begin{equation*}
\nabla \times ( J^T \varphi) = \det J J^{-1} (\nabla' \times \varphi_1). 
\end{equation*}
It follows that, after a change of variables,  
\begin{equation*}
\int_{D} (\nabla \times H) J^T \varphi  = -  \int_{D}   \det J J^{-T} H  (\nabla' \times \varphi_1)  = - \sign \cdot  \int_{D'}   H'  (\nabla' \times \varphi_1), 
\end{equation*}
which implies
\begin{equation}\label{p3}
\int_{D} (\nabla \times H) J^T \varphi  = \sign \cdot  \int_{D'}   (\nabla' \times H') \varphi_1. 
\end{equation}
A combination of \eqref{p1}, \eqref{p2}, and \eqref{p3} yields 
\begin{equation*}
 \int_{D'}   (\nabla' \times H') \varphi_1 =  \int_{D'} (- i k \eps' E' + j') \varphi_1. 
\end{equation*}
Since $\varphi$ is arbitrary, so is $\varphi_1$, we derive that $\nabla' \times H'= - i k \eps' E' + j'$. Identity $\nabla' \times E' = i k \mu' H'$ can be obtained form $\nabla' \times H' = - i k \eps' E' + j'$ by interchanging the role of $E$ and $H$, $\eps$ and $\mu$, replacing $k$ by $-k$, and taking $j=0$.

We next prove \eqref{bdry}. Fix $\varphi \in [C^1(\bar D)]^3$. We have 
\begin{equation*}
\int_{D} (\nabla \times H) J^T \varphi = \int_{D} (- i k \eps E + j) J^T \varphi. 
\end{equation*}
This implies, by an integration by parts,  
\begin{equation*}
\int_{D} H \nabla \times (J^T \varphi) - \int_{\partial D} h (J^T \varphi)  = \int_{D} (- i k \eps E + j) J^T \varphi. 
\end{equation*}
Since \eqref{p2} also holds for $\varphi \in C^1(\bar D)$, we derive  that 
\begin{equation*}
\sign \cdot  \int_{D'} H' (\nabla' \times \varphi_1) - \int_{\partial D} h (J^T \varphi)  = \sign \cdot  \int_{D'} (- i k \eps' E' + j') \varphi_1, 
\end{equation*}
where $\varphi_1(x') = \varphi(x)$.  Integration by parts gives
\begin{equation*}
\sign \cdot \int_{D'} (\nabla' \times H') \varphi_1  + \sign \cdot  \int_{\partial D'} \varphi_1 (H' \times \nu')  - \int_{\partial D} h (J^T \varphi) = \sign \cdot  \int_{D'} (- i k \eps' E' + j') \varphi_1. 
\end{equation*}
We obtain 
\begin{equation*}
\int_{\partial D'} (H' \times \nu') \varphi_1  = \sign \cdot  \int_{\partial D} {\bf J} h \varphi, 
\end{equation*}
where ${\bf J} = \nabla_{\partial D} {\bf T}$ since $h \cdot \nu = 0$ on $\partial D$.  A change of variable yields 
\begin{equation*}
H' \times \nu' = {\bf T}_* h \mbox{ on } \partial D'. 
\end{equation*}
Similarly, we obtain $E' \times \nu' = {\bf T}_* g \mbox{ on } \partial D'$. The proof is complete. 
\proofend

\providecommand{\bysame}{\leavevmode\hbox to3em{\hrulefill}\thinspace}
\providecommand{\MR}{\relax\ifhmode\unskip\space\fi MR }
\providecommand{\MRhref}[2]{%
  \href{http://www.ams.org/mathscinet-getitem?mr=#1}{#2}
}
\providecommand{\href}[2]{#2}


\begin{thebibliography}{10}

  
\bibitem{AlonsoValli}
A.~Alonso and A.~Valli, \emph{{Some remarks on the characterization of the
  space of tangential traces of $H(\mbox{rot}; \Omega)$ and the construction of
  an extension operator}}, Manuscripta Math. \textbf{89} (1996), 159--178.

\bibitem{AmmariCiraoloKangLeeMilton}
H.~Ammari, G.~Ciraolo, H.~Kang, H.~Lee, and G.~W. Milton,   \emph{{Spectral theory of a Neumann-Poincar\'e-type operator and
  analysis of cloaking due to anomalous localized resonance}}, Arch. Rational
  Mech. Anal. \textbf{218} (2013), 667--692.



\bibitem{BallCapdeboscq}
J.~Ball, Y.~Capdeboscq, and B.~Tsering-Xiao, \emph{{On uniqueness for time
  harmonic anisotropic Maxwell's equations with piecewise regular
  coefficients}}, Math. Models Methods Appl. Sci. \textbf{22} (2012), 1250036.


\bibitem{BouchitteSchweizer10}
G.~Bouchitt\'e and B.~Schweizer, \emph{{Cloaking of small objects by anomalous
  localized resonance}}, Quart. J. Mech. Appl. Math. \textbf{63} (2010),
  437--463.

\bibitem{BCostabel}
A.~Buffa, M.~Costabel, and D.~Sheen, \emph{{On traces for $H(\curl,\Omega)$ in
  Lipschitz domains}}, J. Math. Anal. Appl. \textbf{276} (2002), 845--867.


\bibitem{ColtonKressInverse}
D.~Colton and R.~Kress, \emph{{Inverse acoustic and electromagnetic scattering
  theory}}, second ed., Applied Mathematical Sciences, vol.~98,
  Springer-Verlag, Berlin, 1998.
  
\bibitem{Costabel}
M.~Costabel, \emph{{A remark on the regularity of solutions of Maxwell's
  equations on Lipschitz domains}}, Math. Methods Appl. Sci. \textbf{12}
  (1990), 365--368.




\bibitem{HaddarJolyNguyen2}
H.~Haddar, P.~Joly, and H-M. Nguyen, \emph{{Generalized impedance boundary
  conditions for scattering problems from strongly absorbing obstacles: the
  case of Maxwell's equations}}, Math. Models Methods Appl. Sci. \textbf{18}
  (2008), 1787--1827.

\bibitem{KohnLu}
R.~Kohn, J.~Lu, B.~Schweizer, and M.~I. Weinstein, \emph{A variational
  perspective on cloaking by anomalous localized resonance}, Comm. Math. Phys.
  \textbf{328} (2014), 1--27.

\bibitem{LaiChenZhangChanComplementary}
Y.~Lai, H.~Chen, Z.~Zhang, and C.~T. Chan, \emph{{Complementary media
  invisibility cloak that cloaks objects at a distance outside the cloaking
  shell}}, Phys. Rev. Lett. \textbf{102} (2009).

\bibitem{Leis}
R.~Leis, \emph{{Initial-boundary value problems in mathematical physics}}, B.
  G. Teubner, Stuttgart; John Wiley-Sons, Ltd., Chichester, 1986.


\bibitem{MiltonNicorovici}
G.~W. Milton and N.~A. Nicorovici, \emph{{On the cloaking effects associated
  with anomalous localized resonance}}, Proc. R. Soc. Lond. Ser. A \textbf{462}
  (2006), 3027--3059.


\bibitem{MiltonNicoroviciMcPhedranPodolskiy}
G.~W. Milton, N.~A. Nicorovici, R.~C. McPhedran, and V.~A. Podolskiy, \emph{{A
  proof of superlensing in the quasistatic regime and limitations of
  superlenses in this regime due to anomalous localized resonance }}, Proc. R.
  Soc. Lond. Ser. A \textbf{461} (2005), 3999--4034.


\bibitem{Monk} P.~Monk,  \emph{Finite element methods for Maxwell's equations}. Numerical Mathematics and Scientific Computation. Oxford University Press, New York, 2003.



\bibitem{Ng-Complementary}
H-M. Nguyen, \emph{{Asymptotic behavior of solutions to the Helmholtz
  equations with sign changing coefficients}}, Trans. Amer. Math. Soc., {\bf 367} (2015), 6581--6595. 

\bibitem{Ng-Superlensing}
H-M. Nguyen, \emph{{Superlensing using complementary media}}, Ann. Inst. H.
  Poincar\'e Anal. Non Lin\'eaire,  \textbf{32} (2015), 471--484.


\bibitem{Ng-CALR-CRAS}
H-M. Nguyen, \emph{{Cloaking via anomalous localized resonance. A connection
  between the localized resonance and the blow up of the power for doubly
  complementary media}}, C. R. Math. Acad. Sci. Paris, \textbf{353} (2015),
  41--46.

\bibitem{Ng-CALR}
H-M. Nguyen, \emph{{Cloaking via anomalous localized resonance for doubly
  complementary media in the quasi static regime}}, J. Eur. Math. Soc. (JEMS) {\bf 17} (2015), 1327--1365. 



\bibitem{Ng-Negative-cloaking}
H-M. Nguyen, \emph{{Cloaking using complementary media in the quasistatic regime}}, Ann. Inst. H.
  Poincar\'e Anal. Non Lin\'eaire, doi: 10.1016/j.anihpc.2015.06.004, 
  http://arxiv.org/pdf/1310.5483.pdf.


\bibitem{Ng-WP}
H-M. Nguyen, \emph{{Limiting absorption principle and well-posedness for the
Helmholtz equation with sign changing coefficients}}, J. Math. Pures Appl., doi:10.1016/j.matpur.2016.02.013,  http://arxiv.org/pdf/1507.01730v2.pdf. 

\bibitem{Ng-Negative-Review} H-M.Nguyen, \emph{{Negative index materials and their applications: recent mathematics progress}}, 
Chin.  Ann. Math., to appear, http://cama.epfl.ch/files/content/sites/cama/files\\/documents/Survey.pdf. 



\bibitem{Ng-CALR-finite} H-M. Nguyen, \emph{{Cloaking via anomalous localized resonance for doubly
complementary media in the finite frequency regime}}, submitted, http://arxiv.org/abs/1511.08053. 


\bibitem{MinhLoc1} 
H-M. Nguyen and H.~L. Nguyen, \emph{Complete resonance and localized resonance in plasmonic
  structures}, ESAIM: Math. Model. Numer. Anal. \textbf{49} (2015), 741--754.
  	
	
\bibitem{MinhLoc2}
H-M. Nguyen and H.~L. Nguyen, \emph{A three sphere inequality for the Helmholtz equation and its
  application to cloaking using complementary in the finite frequency regime}, Trans. Amer. Math. Soc. B, {\bf 2} (2015), 93--112.  
  
\bibitem{Tu}
T.~Nguyen and J-N. Wang, \emph{{Quantitative uniqueness estimate for the
  Maxwell system with Lipschitz anisotropic media}}, Proc. Amer. Math. Soc.
  \textbf{140} (2012), 595--605.


\bibitem{NicoroviciMcPhedranMilton94}
N.~A. Nicorovici, R.~C. McPhedran, and G.~W. Milton, \emph{{Optical and
  dielectric properties of partially resonant composites}}, Phys. Rev. B
  \textbf{49} (1994), 8479--8482.

\bibitem{MiltonNicorovici}
G.~W. Milton and N. A. Nicorovici, \emph{{On the cloaking effects associated
  with anomalous localized resonance}}, Proc. R. Soc. Lond. Ser. A \textbf{462}
  (2006), 3027--3059.





\bibitem{Paquet}
L.~Paquet, \emph{{Probl\`emes mixtes pour le syst\`eme de Maxwell}}, Ann. Fac.
 Sci. Toulouse Math. \textbf{4} (1982), 103--141.


\bibitem{PendryNegative}
J.~B. Pendry, \emph{{Negative refraction makes a perfect lens}}, Phys. Rev.
  Lett. \textbf{85} (2000), 3966--3969.

\bibitem{PendryCylindricalLenses}
J.~B. Pendry, \emph{{Perfect cylindrical lenses}}, Optics Express \textbf{1} (2003),
  755--760.


\bibitem{PendryRamakrishna-1}
S.~A. Ramakrishna and J.~B. Pendry, \emph{Focusing light using negative
  refraction}, J. Phys.: Condens. Matter \textbf{14} (2002), 8463-8479.

\bibitem{PendryRamakrishna}
S.~A. Ramakrishna and J.~B. Pendry, \emph{{Spherical perfect lens: Solutions of
  Maxwell's equations for spherical geometry}}, Phys. Rev. B \textbf{69}
  (2004), 115115.
  
\bibitem{Protter60}
M.~H. Protter, \emph{{Unique continuation for elliptic equations}}, Trans.
Amer. Math. Soc. \textbf{95} (1960), 81--91.

\bibitem{ShelbySmithSchultz}
R.~A. Shelby, D.~R. Smith, and S.~Schultz, \emph{{Experimental Verification of
  a Negative Index of Refraction}}, Science \textbf{292} (2001), 77--79.


\bibitem{Veselago}
V.~G. Veselago, \emph{{The electrodynamics of substances with simultaneously
  negative values of $\eps$ and $\mu$}}, Usp. Fiz. Nauk \textbf{92} (1964),
  517--526.


\bibitem{Weber} Ch. Weber, \emph{A local compactness theorem for Maxwell's equations}, 
Math. Methods Appl. Sci. {\bf 2} (1980),  12--25. 

\end{thebibliography}
\end{document}